\documentclass{emulateapj}

\newcommand{\msun}{\,\hbox{$M_{\odot}$}}
\newcommand{\lsun}{\,\hbox{$L_{\odot}$}}

\newcommand{\lir}{\,\hbox{$L_{\rm IR}$}}
\newcommand{\spi}{{\it Spitzer}}

\newcommand{\um}{\,\hbox{$\mu$m}}

\newcommand{\neii}{\,\hbox{[\ion{Ne}{2}]}}
\newcommand{\neiii}{\,\hbox{[\ion{Ne}{3}]}}
\newcommand{\nev}{\,\hbox{[\ion{Ne}{5}]}}
\newcommand{\sitwo}{\,\hbox{[\ion{Si}{2}]}}

\newcommand{\oiv}{\,\hbox{[\ion{O}{4}]}}
\newcommand{\oi}{\,\hbox{[\ion{O}{1}]}}
\newcommand{\oiii}{\,\hbox{[\ion{O}{3}]}}
\newcommand{\cii}{\,\hbox{[\ion{C}{2}]}}

\newcommand{\htwo}{\,\hbox{$\rm{H_ 2}$}}

\shorttitle{FIR spectroscopic cosmological surveys}
\shortauthors{Spinoglio et al.}

\usepackage{graphicx}
\usepackage{lscape}
\usepackage{amsmath}
\usepackage{lastpage}
\usepackage{longtable}

\begin{document}

\DeclareGraphicsExtensions{.pdf,.gif,.jpg}

\title{Far-IR/submillimeter Spectroscopic Cosmological Surveys: \\
predictions of infrared line luminosity functions for {\MakeLowercase z}$<$4 galaxies
}

\author{Luigi Spinoglio\altaffilmark{1}, Kalliopi M. Dasyra\altaffilmark{2,3}, Alberto Franceschini\altaffilmark{4},
Carlotta Gruppioni\altaffilmark{5},  Elisabetta Valiante\altaffilmark{6} and 
Kate Isaak\altaffilmark{7}}
\altaffiltext{1}{Istituto di Fisica dello Spazio Interplanetario, INAF, Via Fosso del Cavaliere 100, I-00133 Roma, Italy}
\email{luigi.spinoglio@ifsi-roma.inaf.it}
\altaffiltext{2}{Laboratoire AIM, CEA/DSM-CNRS-Universit\'e Paris Diderot, Irfu/Service dÕAstrophysique, CEA Saclay, F-91191 Gif-sur-Yvette, France}
\altaffiltext{3}{Observatoire de Paris, LERMA (CNRS:UMR8112), 61 Av. de l\'\ Observatoire, F-75014, Paris, France }
\altaffiltext{4}{Dipartimento di Astronomia - Universit\'a di Padova, Vicolo dell\'\ Osservatorio 5, 35122 Padova, Italy}
\altaffiltext{5}{Osservatorio Astronomico di Bologna - INAF, Via Ranzani 1, 40127, Bologna, Italy}
\altaffiltext{6}{Department of Physics and Astronomy, University of British Columbia, 6224 Agricultural Road, Vancouver, BC V6T 1Z1}
\altaffiltext{7}{ESA Research \& Scientific Support Department - ESTEC, Keplerlaan 1, 2200 AG Noordwijk, The Netherlands}

\clearpage

\begin{abstract}
Star formation and accretion onto supermassive black holes in the nuclei of galaxies are the two most energetic processes in the Universe, 
producing the bulk of the observed emission throughout its history. We simulated the luminosity functions of star-forming and active galaxies for 
spectral lines that are thought to be good spectroscopic tracers of either phenomenon, as a function of redshift. We focused on the infrared (IR) 
and sub-millimeter domains, where the effects of dust obscuration are minimal. Using three different and independent theoretical models for 
galaxy formation and evolution, constrained by  multi-wavelength luminosity functions, we computed the number of  star-forming 
and active galaxies per IR luminosity and redshift bin. We converted the continuum luminosity counts into spectral line counts using relationships 
that we calibrated on mid- and far-IR spectroscopic surveys of galaxies in the local universe. Our results demonstrate that future facilities optimized 
for survey-mode observations, i.e., the Space Infrared Telescope for Cosmology and Astrophysics (SPICA) and the Cerro Chajnantor  
Atacama Telescope (CCAT), will be able to observe thousands of $z$$>$1 galaxies in key fine-structure lines, e.g., \sitwo, \oi, \oiii, \cii , 
in a half-square-degree survey, with one hour integration time per field of view.
Fainter lines such as \oiv, \nev\ and \htwo\ (0-0)S1 will be observed in several tens of  bright galaxies at $1<z<2$, while diagnostic diagrams of 
active-nucleus vs star-formation activity will be feasible even for normal $z$$\sim$1 galaxies. We discuss the new parameter space that these future
telescopes will cover and that strongly motivate their construction.
\end{abstract}


\section{Introduction}
\label{intro}

Tremendous progress in infrared (IR) astronomy was made in the last decade, largely driven by the launch of the \spi\ and {\it Herschel} 
space telescopes. Nonetheless, several basic questions remain unanswered in the fields of observational cosmology and galaxy formation 
and evolution, emphasizing the need for future missions. These regard our incomplete knowledge on how the primordial gas collapses to
form new stars, on the different modes of star formation, and on the potentially coeval growth of black holes and galaxies. Below, we elaborate 
on each of these turn. We argue that fine-structure and molecular lines observable in IR lines can help us address them, and we present a set
of simulations that support the need for future survey-oriented facilities that can make a strong impact on studies of galaxy evolution.

The collapse of the primordial gas at very high redshifts that led to the formation of the 
first stars and galaxies is thought to have occured via \htwo\ line emission, which acts as a very effective cooling mechanism in low-metallicity and low-
temperature ($<$10$^4$ K) environments \citep{wise07,obr09}. Even though \htwo\ remains a main coolant soon after the epoch of reionization,  the 
direct detection of \htwo\ gas at very high $z$ is yet to be achieved observational point of view. 

Recent studies of the global infrared continuum 
and molecular gas properties of galaxies in  the local and intermediate/high redshift Universe suggest that mergers and non- or weakly-interacting 
star-forming galaxies follow two separate Kennicutt-Schmidt relations with similar exponents, but different normalizations \citep{Genzel2010,Daddi2010,
Gracia-Carpio2011}. At high-$z$, both galaxy mergers and accretion of cold gas via cooling flows have been suggested as sufficient mechanisms to 
produce IR luminosities $>$10$^{12}$ \lsun\ \citep{Powell2011}, unlike in the local Universe where ultraluminous IR galaxies (ULIRGs) are predominantly 
associated with mergers of comparable mass spirals \citep{Dasyra2006}. The parameters that determine the mode of star formation and control 
its efficiency are not yet understood.  While {\it Spitzer} and {\it Herschel} helped us to identify and determine the star-formation rates of ULIRGs 
at \textit{z}$\lesssim$3, the bulk of star formation at these redshifts is thought to take place in galaxies of lower luminosity \citep{per05,Reddy2010,ly11}. 
Characterizing the star formation in such systems is essential for understanding the formation of present-day ellipticals and downsizing \citep{Cowie1996}. 

Studies of local massive galaxies and theoretical models suggest that most spheroids host massive black holes \citep{Richstone1998,Fabian1999}. 
The observed correlations between the masses of these black holes and the luminosities and stellar velocity dispersions of their host galaxies 
\citep{kor92, Mag98, Fer00, tre02, Fer05, Sha09, gul09} are remarkable, given the vastly different scales that they involve. The enormous difference 
between the black hole Schwarzschild radius and the characteristic radius of the bulge indicates that these relations possibly reflect the coeval 
formation of the two in a common gravitational potential. To date, this scenario has been primarily tested through the comparison of the star-formation 
rate vs the black hole accretion rate as a function of look-back time \citep{Treu04,Mar04,Mer04, shim09,gru11}. Indeed, both activities have been found to 
peak at comparable redshifts, between $z\sim$ 1-3.  To further unravel and understand the relationship between black hole growth and bulge formation we 
need to compare the shape of the mass functions of galaxies with those of actively accreting black holes \citep{Dasyra2011}. Mass estimates 
of black holes in obscured systems will be needed for this purpose.

The complex relationship between star formation and active galactic nucleus (AGN) activity, which could extend over long duty cycles 
(of a few hundreds of Myrs), constrains the quantities that can be used as pure diagnostics of either activity. Present cosmological surveys are 
hampered when it comes to disentangling AGN from starburst activity. For example, an unambiguous AGN indicator in the IR is the continuum 
emission from dust just below sublimation temperature in the AGN torus \citep{Schweitzer2006, Mor2009, Mullaney2010}. However, its 
detection is often uncertain. The use of a torus model to reproduce the observed spectral energy distribution (SED) of galaxies can depend on 
the choice of the star-forming galaxy template that is simultaneously used for the same purpose. Selecting high-$z$ sources with unambiguous evidence for 
hot dust--that is independent of the chosen star-forming galaxy template--can lead to small, biased samples of luminous AGN. Less strict criteria can lead to 
incomplete AGN samples that are contaminated by starburst galaxies. This is frequently the case for AGN selected based on mid-IR color-color diagrams 
\citep {la04, ste05, Barmby2006}. The other unambiguous signature of a hard, AGN-related radiation field is the line emission from ionic species 
that require $\gtrsim$100eV for their creation. A line often used for this purpose is [NeV] at 14.32\um , which however was only seen in stacked spectra 
of $z\gtrsim$1 galaxies with \spi\ \citep{Dasyra2009}.

Observing new FIR-selected galaxy samples, or sampling the peak of the IR SEDs of presently known samples, will not only enable us to address such open 
questions, but to also obtain a more coherent view of high-$z$ galaxy populations. The thousands of 24\um\ -selected, IR-bright galaxies discovered by \spi\ 
\citep{Papovich2004, Rigby2004, Fadda2004, LeFloc'h2004, Houck2005, Huang2009}, the hundreds of mm galaxies \citep[e.g. ][]{Smail1997, Hughes1998, Barger1998, Scott2002, Borys2003, Webb2003, Coppin2006}, 
\citep[e.g. ][]{Greve2004, Laurent2005, Scott2008, Perera2008, Austermann2010},
the tens of thousands far-IR and sub-mm galaxies detected at high-redshifts by  the {\it Herschel} observatory \citep[e.g. ][]{ber10, oli10}, 
and the thousands of near IR dropouts 
\citep[e.g. ][]{Caputi2005, Daddi2005, Papovich2006} known to date are often treated as physically disconnected populations. 

Spectroscopic surveys with future missions or facilities will help us address the above-mentioned points. Expected to be at least one order 
of magnitude more sensitive than their predecessors, future spectrographs will observe ionic and molecular gas lines, in addition to the dust broadband features 
and their underlying continua, for large galaxy samples over wide fields of view. In the spirit of demonstrating what can be done in the future  in the IR/submm 
domain in the field of galaxy evolution,  we present simulations of the number of galaxies that will be observed in  
key IR lines and features  at each $z$ range with the SPace IR telescope for Cosmology and Astrophysics (SPICA) and the Cerro Chajnantor 
Atacama Telescope (CCAT). We focus on SPICA and CCAT as these 
are conceived to be large blind survey machines, unlike the James Webb Space Telescope (JWST) and the Atacama Large Millimeter Array (ALMA), that are 
optimized for deep single-source observations. 

The paper is organized as follows. In Section~\ref{sec:simulations} we present the simulations, and the physical assumptions that go into them, including the 
adopted line-to-bolometric luminosity conversion functions that we measured from observations of local galaxies. 
The main results of the simulations, i.e., the number of galaxies 
and AGN that can be detected for each ionized or molecular gas line with $z$ for a given flux limit, are presented in Section~\ref{sec:results}, followed by a
discussion in Section~\ref{sec:discussion}. Throughout the paper we will adopt the standard cosmological model  
with $\Omega_{M}$= 0.27, $\Omega_{\Lambda}$=0.73, H$_{0}$=71 km s$^{-1}$ Mpc$^{-1}$.

\section{Simulating spectroscopic surveys in the IR}
\label{sec:simulations}

The computation of the number of objects that future missions will detect in each line was a two-fold process. First, we used 
three galaxy evolution models that are based on galaxy counts and luminosity functions in several bands of the \textit{Spitzer}, 
AKARI and \textit{Herschel} missions to predict the integrated 8-1000\um\ IR luminosity function at different $z$ bins.  
We have considered three different models with the aim of basing our conclusions on solid results, and at the same time
quantifying the range of the predictions. We then 
computed correlations between the IR luminosity and the luminosities of various fine-structure lines,  molecular lines, and dust 
emission features, using samples of local galaxies with complete mid- and far-IR spectroscopic coverage. This enabled us to 
transform the continuum luminosity functions into line luminosity functions up to $z$$<$4 and compute the line detection rates.

  \begin{figure}[ht!]
  \includegraphics[width=\columnwidth]{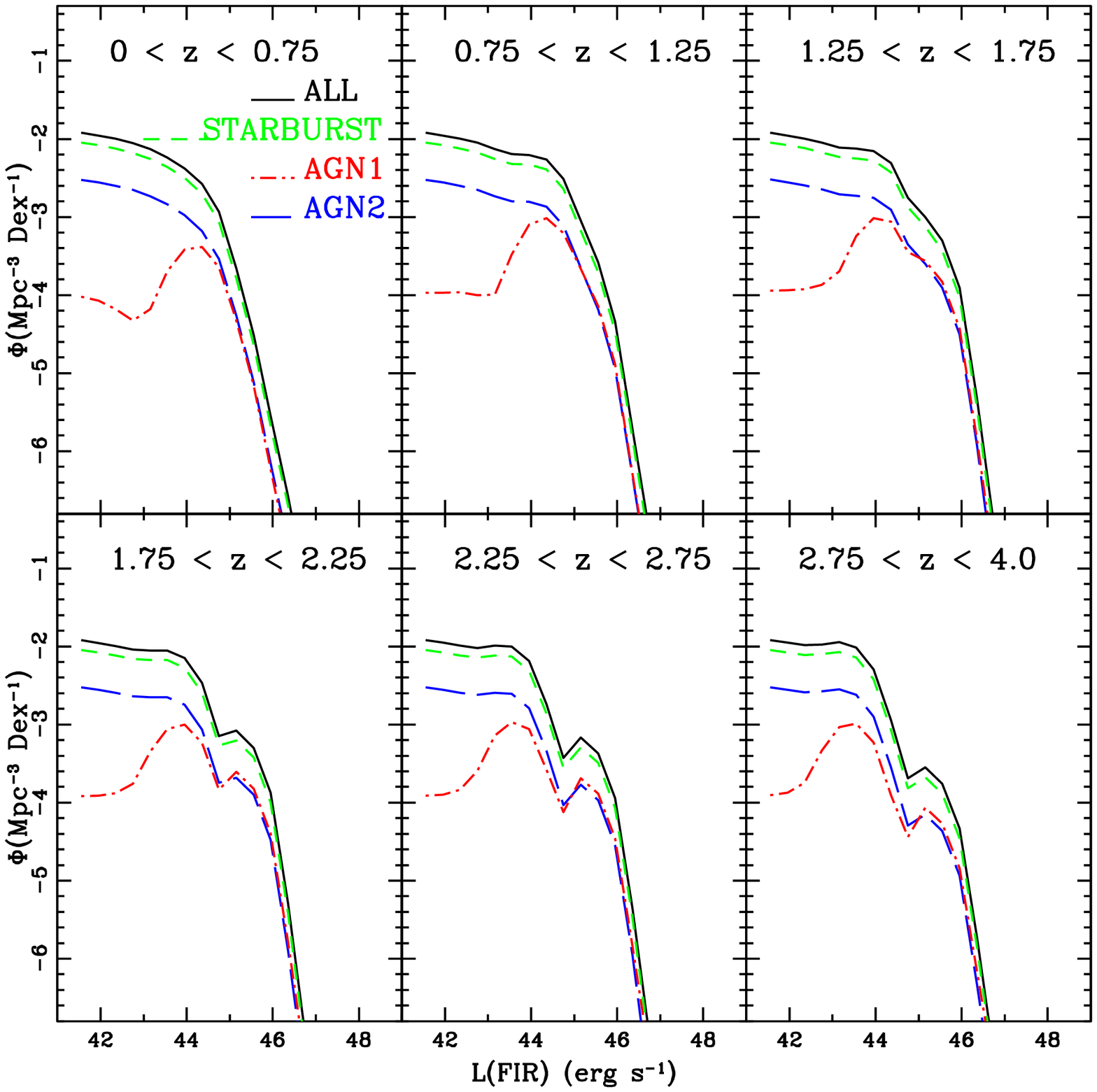}
  \caption{IR (8$-$1000\um ) continuum luminosity function constructed from the \citet{fra10} model.}
  \label{fig:inputlf-fran}
\end{figure}

\subsection{Predicting continuum luminosity functions with the \citet{fra10} model}

As a first approach, we adopted the model developed by \citet{fra10}. This is a backward evolution model that fits essentially all available data from \textit{Spitzer}, ISO, COBE, and SCUBA. Moreover, it includes constraints from preliminary results of the \textit{Herschel Space Observatory} surveys 
\citep[presented in the Science Demonstration Phase papers of][]{oli10, ber10, gle10, Gru10, ngu10}.

The model accounts separately for normal spiral galaxies, actively star-forming galaxies, and for type 1 and 2 AGNs. The actively star-forming population is further split into two galaxy 
classes, i.e.,  moderate-luminosity luminous infrared galaxies (LIRGs) and high-luminosity ULIRGs. The LIRGs have typical IR luminosities $\simeq 10^{11}\ L_\odot$ and the ULIRGs $\simeq 10^{12}\ L_\odot$.  
They are treated as two galaxy classes, with different luminosity functions, evolution rates, and spatial clustering properties. 

The broad-line AGNs--type 1 Seyferts and quasars--were modelled adopting luminosity functions and evolution rates consistent with those observed in optical and mid-IR surveys \citep[e.g.][]{rms,spi95}.
As detailed in \citet{fra10}, important constraints on the infrared evolution properties of type 1 AGNs have been inferred from a flux-limited sample of ${24\mu m}$-selected sources with complete spectroscopic classification, as reported by \citet{rod10}. These objects are easily identified by their flat spectral shapes over the optical through IR wavelength range \citep{sam02}.

Narrow-line type 2 AGNs are instead much more difficult to disentangle from starburst galaxies. Therefore their statistical properties and incidence among the IR population at high redshifts are still essentially unknown. At the present stage of knowledge, the model simply considers type 2 AGNs as a fraction of both low-luminosity and high-luminosity starbursts. In an attempt to constrain such fractions, a follow-up analysis by Franceschini et al. (2011, in preparation) has compared the star-formation and stellar-mass assembly histories of galaxies, and found that the two are consistent with each other if 10\% of the LIRG and 30\% of the ULIRG objects are dominated by obscured (type 2) AGN accretion. These are the fractional contributions for type 2 AGNs adopted in the present work. 
It is clear from the above that this modelling might easily give results inconsistent with the value derived from both predictions by the AGN unification scheme
and observations \citep[e.g., from the SLOAN and FIRST surveys,][]{lu2010} of a $\sim$1 to 3 ratio of type-1 to type 2 objects. Indeed, the validity of the AGN unification at high redshifts has never been proven in detail, and will be one of the major outcomes of futures spectroscopic surveys in the IR. 

Figure~\ref{fig:inputlf-fran} shows the continuum far-IR luminosity functions 
for all populations in each $z$ range considered: 0.$<$z$<$0.75, 0.75$<$z$<$1.25, 
1.25$<$z$<$1.75, 1.75$<$z$<$2.25, 2.25$<$z$<$2.75 and 2.75$<$z$<$4.0.

 \begin{figure}[!ht]
     \includegraphics[width=\columnwidth]{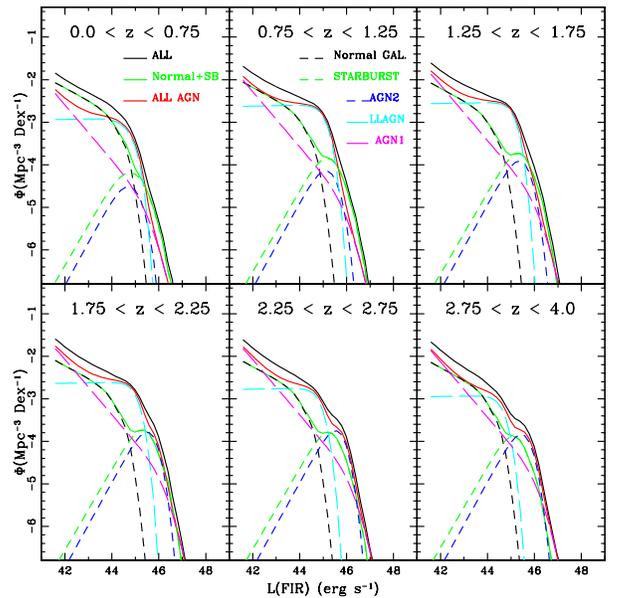}
     \caption{IR luminosity function from \citet{gru11} for various galaxy populations,
   i.e., normal and starburst galaxies, low luminosity AGN, type 1 AGN and type 2 AGN. }
     \label{fig:inputlf-grup}
   \end{figure}

\subsection{Predicting continuum luminosity functions with the \citet{gru11} model}

\citet{gru11} have also developed a backward evolution model fitting 
the main constraints provided by IR/sub-mm  surveys in the 
15$\mu$m to 500$\mu$m range. In the mid-IR (MIR), it uses data from both ISO and {\em Spitzer} that are available in the literature, 
at 15$\mu$m from the ELAIS-S1 \citep{Gru02}, 
HDF-N, HDF-S and Marano fields \citep{elb99}, 
ultradeep lensed \citep{met03}, 
Lockman Deep and Shallow \citep{rod04}, 
and at 24$\mu$m from the GOODS \citep{Papovich2004} 
and SWIRE surveys \citep{shu08}. 
In the FIR, we use data from {\it Herschel}, including those of the very recent PACS Evolutionary Probe Survey \citep{ber10, Gru10}, 
the {\em Herschel}-ATLAS Survey \citep{ea10, cle10}, and the {\em Herschel} Multi-tiered Extra-galactic Survey \citep{oli10, vac10}. 
The model uses the classical approach of evolving a  local luminosity function in luminosity and/or density with $z$, with 
a different evolutionary description for galaxies and AGN. What differentiates this model from others is the spectral energy distribution (SED) 
scheme that it uses to distinguish between the different IR-bright galaxy populations. The scheme is based on a large spectroscopic study of 
MIR-selected sources \citep{Gru08}, and it divides sources into five (instead of four) broad SED classes. These are the normal spiral galaxies, 
the starburst galaxies, the obscured type 2 AGNs, the unobscured type 1 AGNs, and the objects containing a low luminosity (LL) AGN. In other
words, the authors of this model have put particular emphasis on determining the AGN contribution at low luminosities at all redshifts. The contribution of the various galaxy populations to the IR luminosity function can be seen in Figure~\ref{fig:inputlf-grup}.


  \begin{figure}[!h]
    \includegraphics[width=\columnwidth]{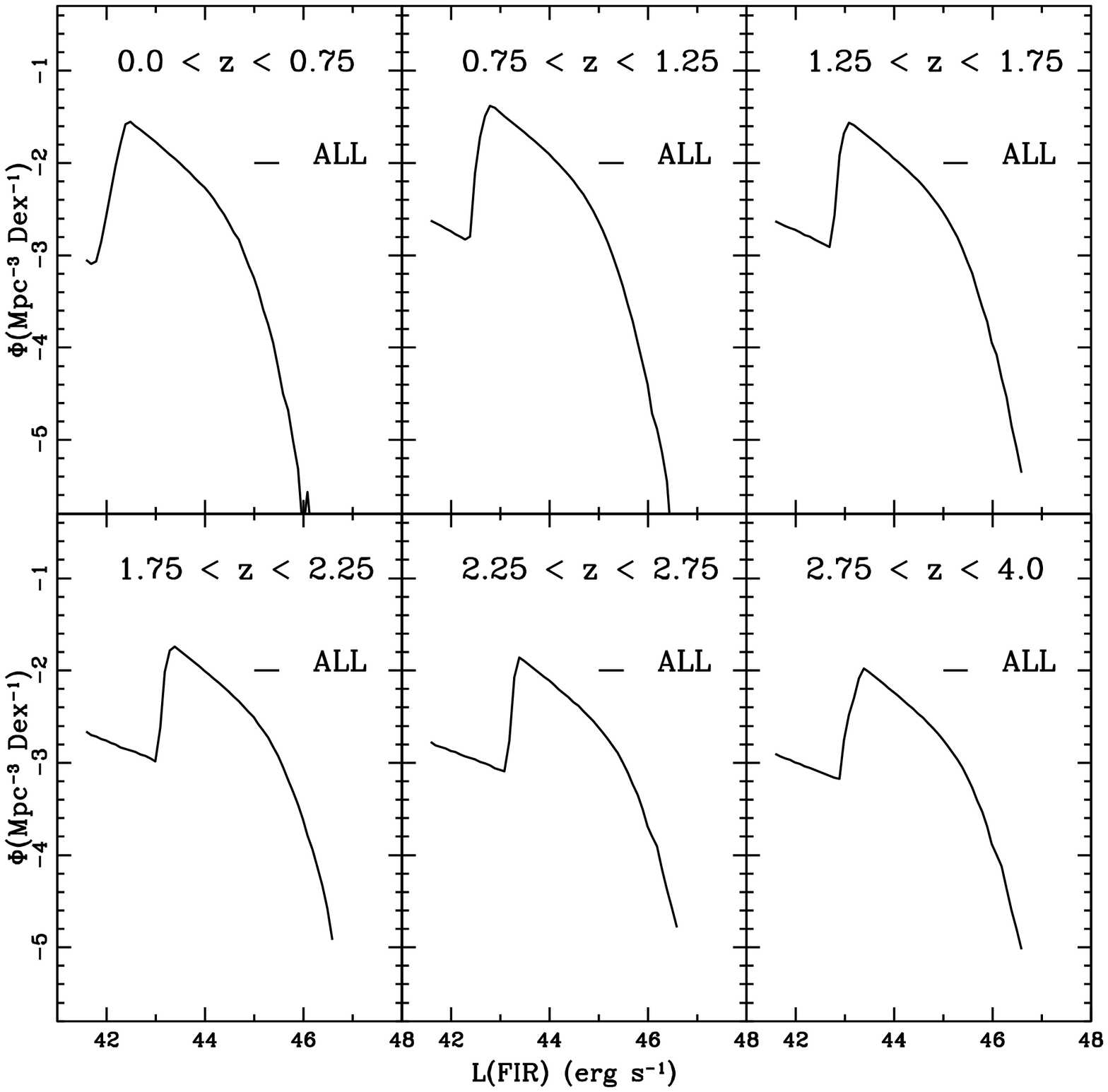}
    \caption{IR luminosity function from \citet{val09}. 
    }
    \label{fig:inputlf-val}
  \end{figure}

\subsection{Predicting continuum luminosity functions with the \citet{val09} model}

The third model that we used is the backward evolution model of \citet{val09}. It was developed using \textit{Spitzer} 
and SCUBA observations, and it has been very successful in predicting \textit{Herschel} results \citep{ber10,alt10,
gle10,oli10}. This model allows us to take into account galaxies that are not 'pure' starburst or 'pure' AGN, and for 
which the ratio between IR lines might not be the ones expected assuming 'pure' SEDs. This is because the model 
considers all infrared galaxies as a single population, assuming that starbursts and AGN coexist. It then uses an 
empirical relation to assign to each galaxy the fraction of the IR luminosity that is powered by the AGN for its given 
luminosity and redshift. This relation was derived using a complete sample of local IRAS galaxies, and extrapolated 
to high $z$ using \textit{Spitzer} and SCUBA observations. Figure~\ref{fig:inputlf-val} shows the IR luminosity function
predicted by this model for each $z$ bin.

\label{}
 \begin{figure}[t]
  \begin{center}
    \includegraphics[width=\columnwidth]{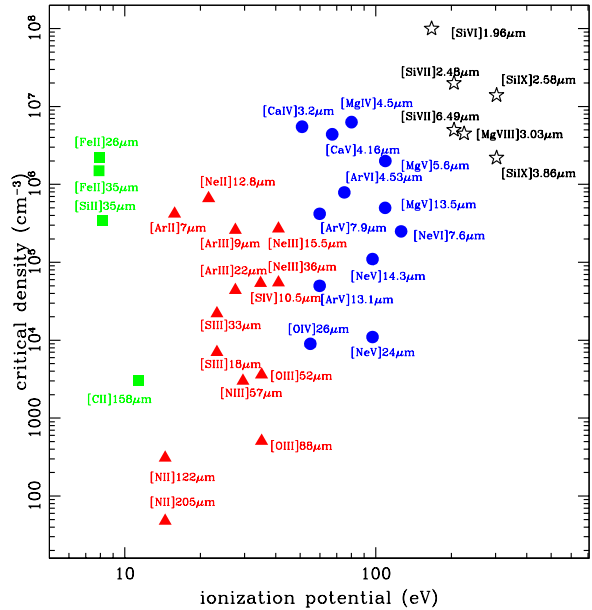}
  \end{center}
  \caption{Critical density for collisional de-excitation vs. ionization potential of IR fine-structure lines. 
  }  
\label{fig:ip_density}
\end{figure}

\subsection{Converting continuum luminosity functions to line or feature luminosity functions.} 

The galaxy number counts per redshift and bolometric IR luminosity bin that are predicted by each model need to be 
converted into line luminosity functions, in order to estimate the number of objects that will be detectable in various 
lines, and to assess whether several of the open questions presented in Section~\ref{intro} can be addressed. For this 
purpose, we derived correlations between line and continuum luminosities, including lines for which such correlations 
were not previously available in the literature, e.g., that of [SiII] for both AGN and star-forming galaxies. 
We examined the PAH feature at 11.25$\mu$m, the purely rotational H$_2$ (0-0)S1 line at 17.03$\mu$m, and the 
[NeII] 12.8$\mu$m, [NeV] 14.3$\mu$m, [NeIII] 15.5$\mu$m, [SIII] 18.7$\mu$m, [NeV] 24.3$\mu$m, [OIV] 26$\mu$m, 
[SiII] 34.8$\mu$m, [OIII] 52$\mu$m, [NIII] 57$\mu$m, [OI] 63$\mu$m, [OIII] 88$\mu$m, [NII] 121.90$\mu$m, 
[OI] 145.52$\mu$m, and [CII] 157.74$\mu$m fine-structure lines. These lines cover a wide parameter space of the critical
density vs. ionization potential diagram (see Fig.~\ref{fig:ip_density}), tracing different astrophysical conditions: from photodissociation 
regions, to stellar/HII regions, to the AGN and coronal line regions \citep{sm92}. This makes the combination of their 
ratios useful for the creation of AGN vs star-formation diagnostic diagrams \citep[e.g. ][]{sm92, gen98, dal06, smi07}. 

For lines at wavelengths shorter than 35$\mu$m, we used the complete, 12$\mu$m-selected sample of local Seyfert galaxies 
\citep{tom08,tom10} and the \citet{bs09} sample of starburst galaxies to calibrate the line luminosities to \lir . These samples 
have been extensively observed in the MIR with the IRS spectrometer \citep{hou04} onboard \textit{Spitzer} \citep{wer04}, and the \textit{Spitzer} spectra 
have been reduced and analysed in a consistent way. For the starburst galaxies, we excluded all objects for which there was 
evidence for the presence of an AGN from the literature or from the detection of [NeV] \citep[see Table 1 of][]{bs09}. For the long- 
wavelength lines, we used the heterogeneous sample of local galaxies compiled by \citet{bra08} containing all observations 
collected by the LWS spectrometer \citep{cle96} onboard ISO \citep{kes96}. The IR luminosities of the galaxies of our sample have been computed 
from the IRAS fluxes, using the formula of $L_{IR}$\footnote{$L_{IR}$ is computed by fitting a single-temperature dust emissivity model 
($\epsilon$ $\propto$ $\nu^{-1}$) 
to the flux in all four IRAS bands, and should be accurate to $\pm$5$\%$ for dust temperatures in the range 25 - 65 K. We notice that the 
IR luminosities, as defined above, are model-dependent, and therefore could introduce some systematics. However these 
do not affect the derived relations, as they are within the given errors. } 
representing the total mid- and far-infrared luminosity \citep{sm96}. All luminosities are in units of 10$^{41}$ erg s$^{-1}$.

\begin{figure*}[!h]
\begin{center}
{\includegraphics[width=\textwidth]{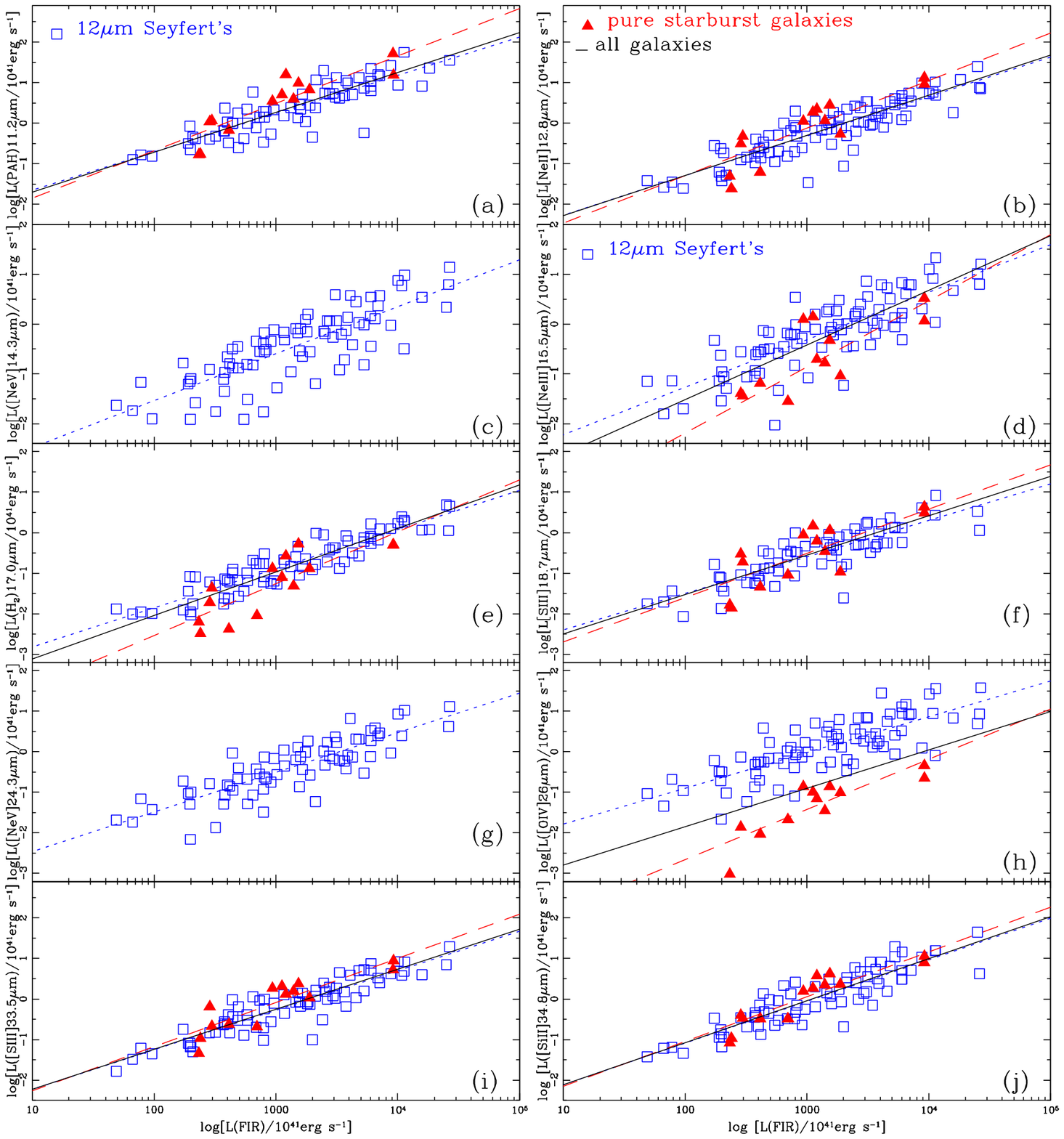}}
\end{center}
\caption{Correlations between the various feature and line luminosities and the far-IR luminosity for the
Seyfert galaxies of the complete 12$\mu$m galaxy sample \citep{tom08, tom10}, and for the pure starburst galaxies 
of the sample of \citet{bs09}. The dotted, broken and solid lines represent the least-squares fit of the data of the Seyfert, 
the pure starburst galaxies and all galaxies populations together, respectively. 
Figures (c) and (g) have only the Seyfert galaxies, because the [NeV] lines
are not detected in starburst galaxies.}
\label{fig:relations_short}
\end{figure*}

Using least-squares fitting, we obtained the following relations for the Seyfert galaxies,
{\footnotesize
\begin{eqnarray}
&& \log(L_{\rm PAH 11.25})=(0.95\pm0.07) \log(L_{IR})-(2.60\pm0.21)\\
&& {\rm with} ~R=0.87, ~n=69, ~\chi^2= 7.1 \nonumber \\
&& \log(L_{\rm [NeII]12.81}) = (0.98\pm0.06) \log(L_{IR}) -(3.25\pm0.18)\\
&& {\rm with} ~R=0.89, ~n=87, ~\chi^2=8.8  \nonumber \\
&& \log(L_{\rm [NeV]14.32}) = (0.94\pm0.08) \log(L_{IR}) - (3.43\pm0.25) \\ 
&& {\rm with} ~R=0.81, ~n=81, ~\chi^2=14.8  \nonumber \\
&& \log(L_{\rm [NeIII]15.55}) = (0.95\pm0.07) \log(L_{IR}) - (3.18\pm0.24) \\
&& {\rm with} ~R=0.82, ~n=87, ~\chi^2=15.0  \nonumber \\
&& \log(L_{\rm (H_{2})17.03}) = (0.97\pm0.05) \log(L_{IR}) - (3.79\pm0.15) \\
&& {\rm with} ~R=0.92, ~n=76, ~\chi^2=5.3  \nonumber \\
&& \log(L_{\rm [SIII]18.71})  = (0.90\pm0.07) \log(L_{IR}) - (3.30\pm0.21) \\
&& {\rm with} ~R=0.85, ~n=70, ~\chi^2=7.7  \nonumber \\
&& \log(L_{\rm [NeV]24.31})   = (0.98\pm0.08) \log(L_{IR}) - (3.45\pm0.24) \\ 
&& {\rm with} ~R=0.84, ~n=71, ~\chi^2=10.5  \nonumber \\
&& \log(L_{[OIV]25.89})   = (0.88\pm0.08) \log(L_{IR}) - (2.66\pm0.25) \\ 
&& {\rm with} ~R=0.79, ~n=83, ~\chi^2=14.9  \nonumber \\
&& \log(L_{\rm [SIII]33.48})  = (0.98\pm0.05) \log(L_{IR}) - (3.21\pm0.17) \\
&& {\rm with} ~R=0.91, ~n=75, ~\chi^2=5.7  \nonumber \\
&& \log(L_{\rm [SiII]34.82})  = (1.03\pm0.06) \log(L_{IR}) - (3.14\pm0.20) \\
&& {\rm with} ~R=0.89, ~n=72, ~\chi^2=7.3 \nonumber 
\end{eqnarray}
}
where for each relation the Pearson coefficient R, the number of considered objects n and the computed $\chi^2$ are given. 
For the starburst galaxies, the corresponding relations are:
{\footnotesize
\begin{eqnarray}
&& \log(L_{\rm PAH 11.25})   = (1.17\pm0.11) \log(L_{IR}) - (3.03\pm0.32) \\
&& {\rm with} ~R=0.95, ~n=14, ~\chi^2=1.4  \nonumber \\
&& \log(L_{\rm [NeII]12.81}) = (1.17\pm0.14) \log(L_{IR}) - (3.65\pm0.40) \\
&& {\rm with} ~R=0.93, ~n=14, ~\chi^2=2.1  \nonumber \\
&& \log(L_{\rm [NeIII]15.55}) = (1.33\pm0.18) \log(L_{IR}) - (4.85\pm0.52) \\
&& {\rm with} ~R=0.90, ~n=15, ~\chi^2=3.9  \nonumber \\
&& \log(L_{\rm (H_{2})17.03}) = (1.28\pm0.14) \log(L_{IR}) - (5.10\pm0.42) \\
&& {\rm with} ~R=0.93, ~n=15, ~\chi^2=2.6  \nonumber \\
&& \log(L_{\rm [SIII]18.71}) = (1.09\pm0.15) \log(L_{IR}) - (3.79\pm0.45) \\
&& {\rm with} ~R=0.89, ~n=15,~\chi^2=3.0  \nonumber \\
&& \log(L_{\rm [OIV]25.89}) = (1.24\pm0.24) \log(L_{IR}) - (5.13\pm0.74) \\ 
&& {\rm with} ~R=0.85, ~n=12, ~\chi^2=1.6  \nonumber \\
&& \log(L_{\rm [SIII]33.48}) = (1.09\pm0.10) \log(L_{IR}) - (3.35\pm0.29) \\ 
&& {\rm with} ~R=0.95, ~n=15, ~\chi^2=1.2  \nonumber \\
&& \log(L_{\rm [SiII]34.82}) = (1.11\pm0.09) \log(L_{IR}) - (3.26\pm0.25) \\ 
&& {\rm with} ~R=0.96, ~n=15, ~\chi^2=0.92  \nonumber
\end{eqnarray}
}
The best-fit solution is shown in Figure~\ref{fig:relations_short} for each of the 
populations. Considering the sum of the populations, we derived the following 
generic relations: 
{\footnotesize
\begin{eqnarray}
&& \log(L_{\rm PAH 11.25})   = (0.98\pm0.06) \log(L_{IR}) - (2.69\pm0.18) \\
&& {\rm with} ~R=0.88, ~n=83 ~\chi^2=9.4  \nonumber \\ 
&& \log(L_{\rm [NeII]12.81}) = (0.99\pm0.06) \log(L_{IR}) - (3.26\pm0.20) \\
&& {\rm with} ~R=0.84, ~n=101, ~\chi^2=17.8  \nonumber \\ 
&& \log(L_{\rm [NeIII]15.55}) = (1.10\pm0.07) \log(L_{IR}) - (3.72\pm0.23) \\
&& {\rm with} ~R=0.83, ~n=102, ~\chi^2=24.2  \nonumber \\ 
&& \log(L_{\rm (H_{2})17.03}) = (1.07\pm0.05) \log(L_{IR}) - (4.19\pm0.16) \\
&& {\rm with} ~R=0.91, ~n=91,  ~\chi^2=10.6  \nonumber \\ 
&& \log(L_{\rm [SIII]18.71}) = (0.97\pm0.06) \log(L_{IR}) - (3.47\pm0.20) \\
&& {\rm with} ~R=0.88, ~n=70, ~\chi^2=7.9  \nonumber \\ 
&& \log(L_{\rm [OIV]25.89}) = (0.95\pm0.11) \log(L_{IR}) - (3.04\pm0.34) \\
&& {\rm with} ~R=0.68, ~n=95, ~\chi^2=36.6  \nonumber \\ 
&& \log(L_{\rm [SIII]33.48}) = (0.99\pm0.05) \log(L_{IR}) - (3.21\pm0.14) \\
&& {\rm with} ~R=0.92, ~n=90, ~\chi^2=7.4  \nonumber \\
&& \log(L_{\rm [SiII]34.82}) = (1.04\pm0.05) \log(L_{IR}) - (3.15\pm0.16) \\
&& {\rm with} ~R=0.91, ~n=87, ~\chi^2=8.4  \nonumber
\end{eqnarray}
}

\begin{figure*} 
\begin{center}
{\includegraphics[width=\textwidth]{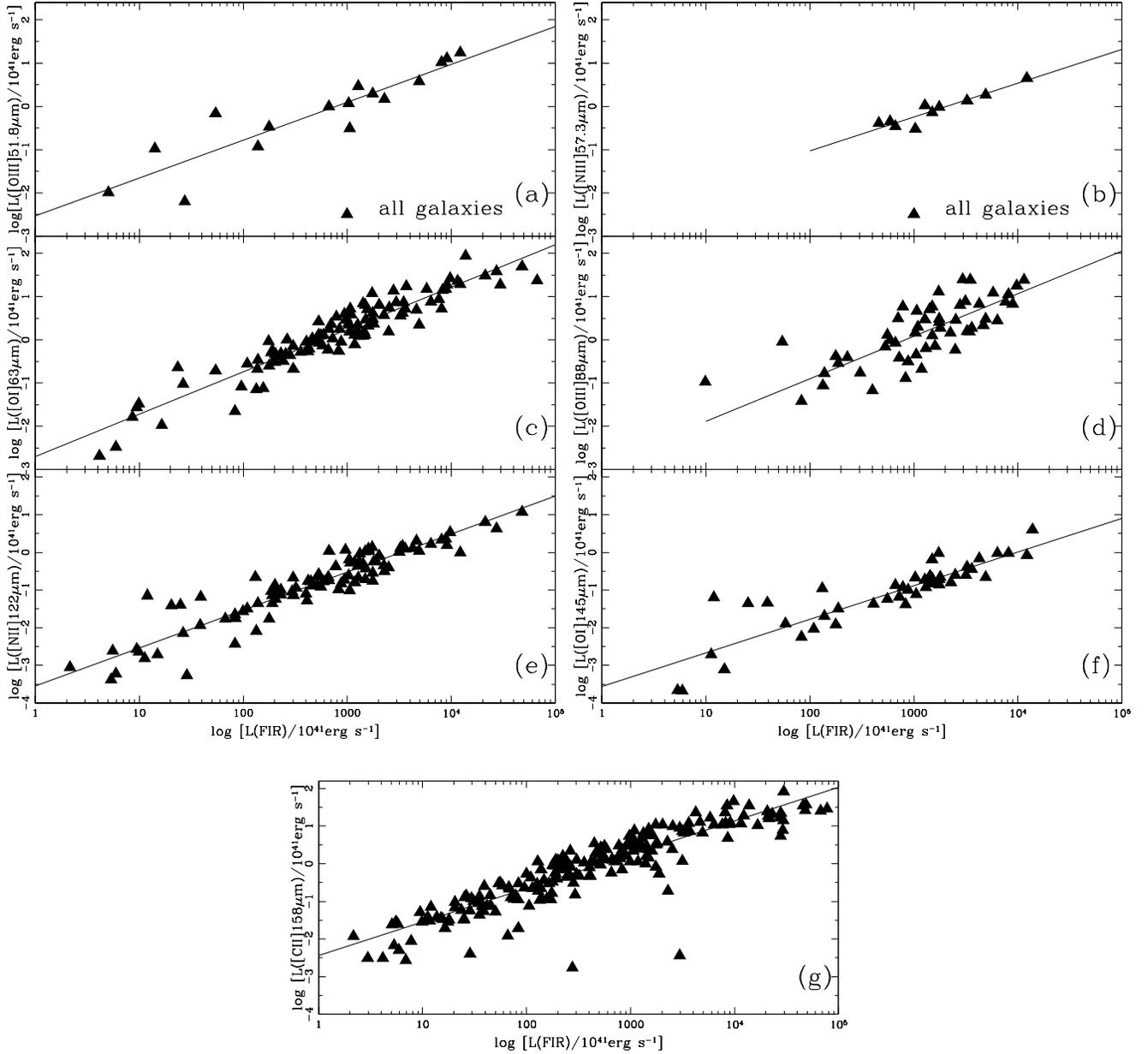}}
\end{center}
\caption{Correlations between the [OIII]52$\mu$m, [NIII]57$\mu$m, [OI]63$\mu$m, [OIII]88$\mu$m, [NII]122$\mu$m, [OI]145$\mu$m and [CII]158$\mu$m luminosity and the far-IR luminosity for the galaxies observed with the ISO-LWS spectrometer \citep{bra08}.}
\label{fig:relations_long}
\end{figure*}

For the far-IR lines (Fig.~\ref{fig:relations_long}), we obtain:
{\footnotesize
\begin{eqnarray}
&& \log(L_{\rm [OIII]51.81}) = (0.88\pm0.10) \log(L_{IR}) - (2.54\pm0.31) \\
&& {\rm with} ~R=0.91, ~n=16, ~\chi^2=2.6  \nonumber \\ 
&& \log(L_{\rm [NIII]57.32}) = (0.78\pm0.10) \log(L_{IR}) - (2.58\pm0.32) \\
&& {\rm with} ~R=0.94,~n=10, ~\chi^2=0.14  \nonumber \\ 
&& \log(L_{\rm [OI]63.18}) = (0.98\pm0.03) \log(L_{IR}) - (2.70\pm0.10) \\
&& {\rm with} ~R=0.94,~n=109, ~\chi^2=9.1  \nonumber\\ 
&& \log(L_{\rm [OIII]88.36}) = (0.98\pm0.10) \log(L_{IR}) - (2.86\pm0.30) \\
&& {\rm with} ~R=0.81,~n=55, ~\chi^2=12.7  \nonumber \\
&& \log(L_{\rm [NII]121.9}) = (1.01\pm0.04) \log(L_{IR}) - (3.54\pm0.11) \\
&& {\rm with} ~R=0.93,~n=100, ~\chi^2=13.3  \nonumber \\
&& \log(L_{\rm [OI]145.5}) = (0.89\pm0.06) \log(L_{IR}) - (3.55\pm0.17) \\
&& {\rm with} ~R=0.91,~n=46, ~\chi^2=10.0  \nonumber \\
&& \log(L_{\rm [CII]157.7}) = (0.89\pm0.03) \log(L_{IR}) - (2.44\pm0.07) \\
&& {\rm with} ~R=0.92,~n=217. ~\chi^2=42.3  \nonumber
\end{eqnarray}
}

For all correlations the hypothesis that the variables are unrelated can be rejected at a level of significance which is always less that 10$^{-3}$.
\citet{wu10} presented the relation between the total infrared luminosity and the PAH emission band at 11.25$\mu$m for
AGN and starburst galaxies of the 24$\mu$m flux limited intermediate redshift ($<z> \sim 0.14$) sample of 5MUSES (Helou et al. 2011, in preparation). Their result
is comparable to ours. By inverting our relations we derive a slope of 1.05$\pm$011 for the Seyfert galaxies and 0.85$\pm$0.08 for the starburst galaxies,
compared to their slopes of 1.00$\pm$0.04 and 0.98$\pm$0.03, respectively.

\section{Results}
\label{sec:results}

\subsection{Number counts per spectral line}

\begin{figure*}[!t]
\includegraphics[width=\textwidth]{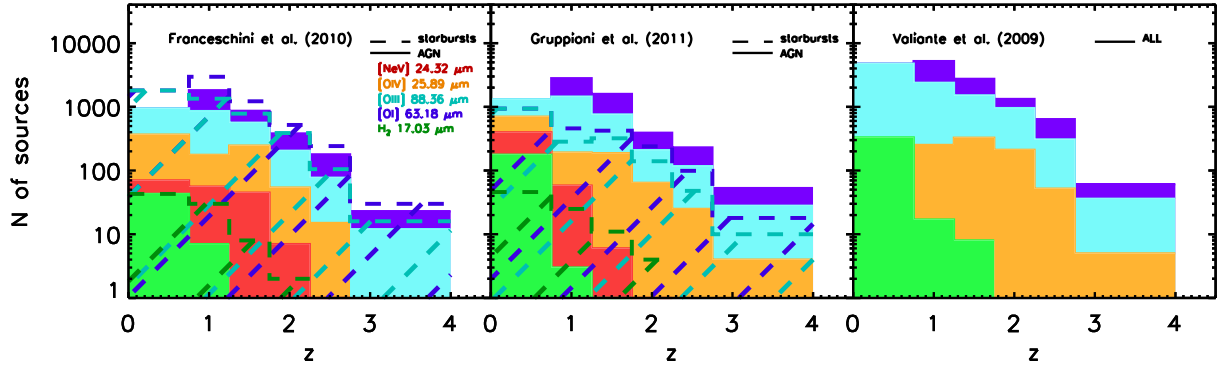}
  \smallskip
   \caption{Number of objects detected per spectral line (and per object type, when applicable) in an hour-long 0.5 deg$^2$ survey with 
   SPICA SAFARI.
}
\label{fig:ncounts_hist}
\end{figure*}

\begin{figure*}[!h]
\includegraphics[width=9truecm]
{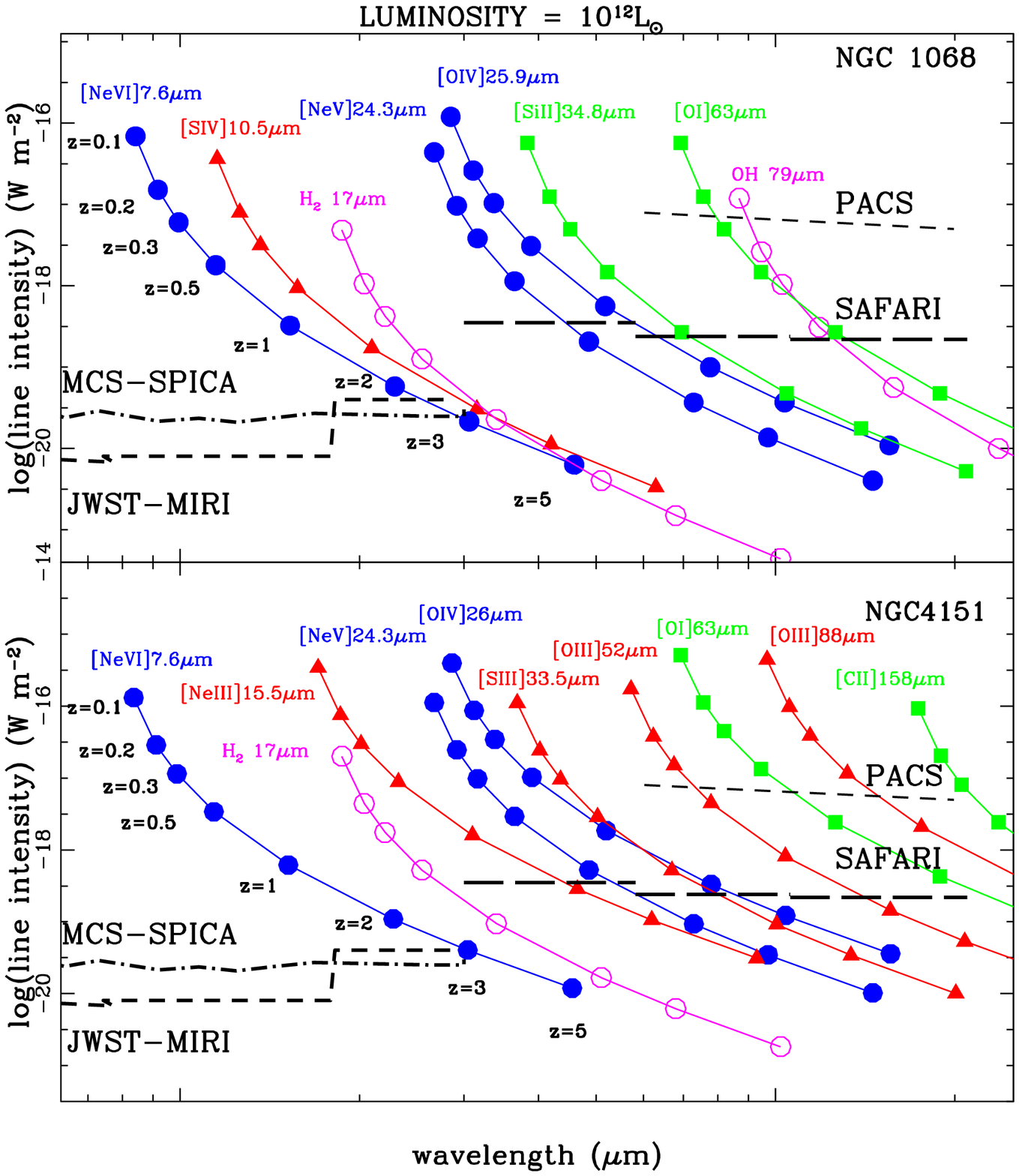} 
\includegraphics[width=9truecm] 
{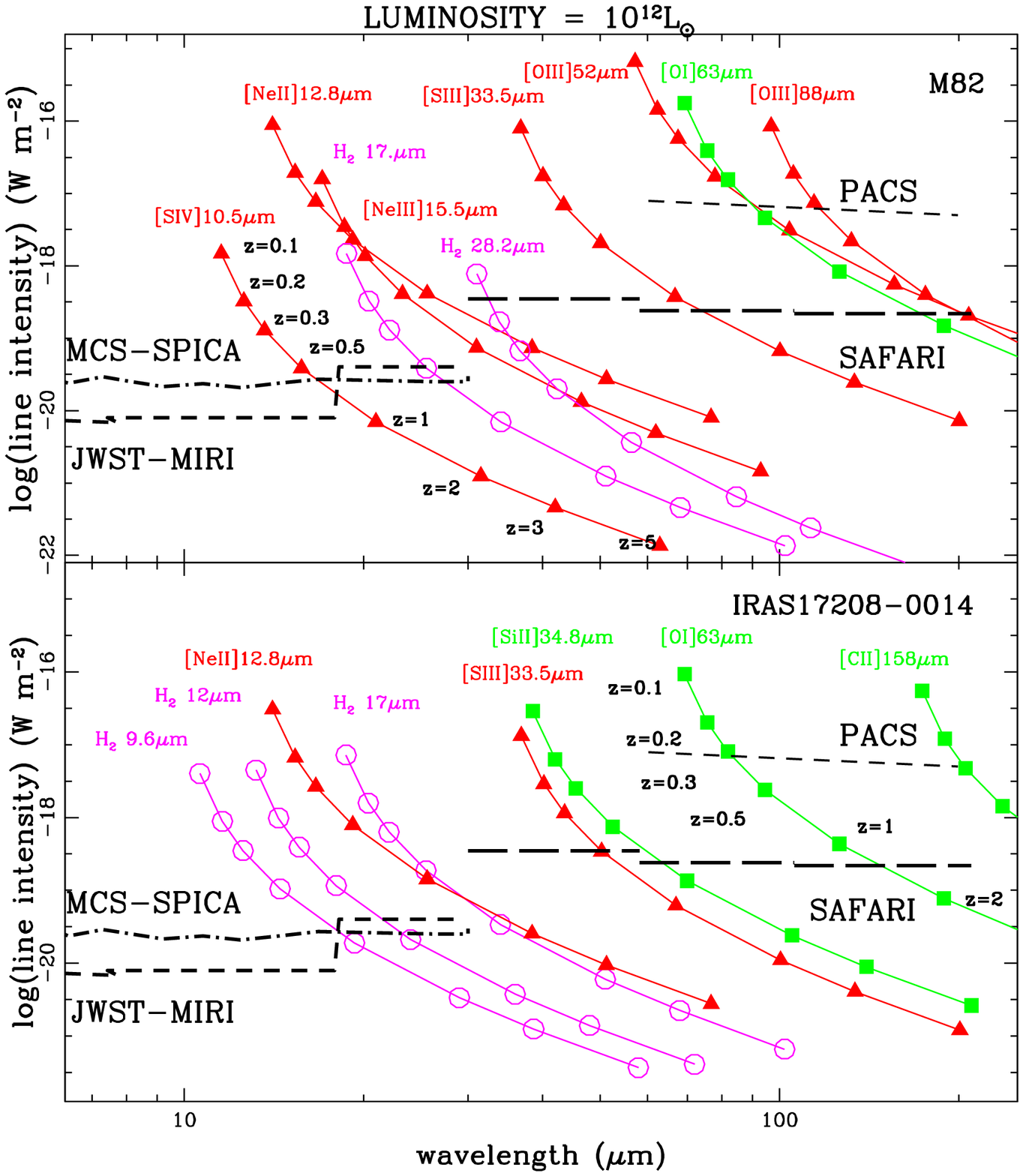} 
\caption{{\bf a)}Predictions of lines observable with SPICA/SAFARI, based on local templates, scaled to an intrinsic luminosity of 10$^{12}$\lsun .  Selected diagnostic lines are shown as a function of $z$ for a type 1 AGN (NGC4151) and a type 2 AGN (NGC1068). 
For comparison, we overplot the 5$\sigma$ (in 1 hour) sensitivity threshold of SPICA-SAFARI as a function of 
wavelength (see Appendix). For completeness, we also report the sensitivities for {\it Herschel} PACS \citep{pog10}, for the Mid-Infrared Instrument \citep[MIRI,][]{wri04}, 
which will be onboard the James Webb Space Telescope \citep{gar06} and which will operate from 5 to 28$\mu$m, and for the Mid-ir Camera and Spectrograph 
(MCS) planned for SPICA \citep{wa10}.
{\bf b)} Same as panel a) for a moderate-luminosity prototypical starburst (M82), and a starburst-dominated ULIRG (IRAS17208-0014).
}
  \label{fig:template_predictions1}
\end{figure*}

To compare and visualize our results for the three galaxy-evolution models, we need to adopt a line-detection sensitivity curve as a function
of wavelength $\lambda$, an integration time, and a field of view for the simulated observations. For this purpose, we opt to use numbers 
relevant to future missions or facilities. For the FIR domain, we use the sensitivity curve proposed for SPICA's far infrared instrument (SAFARI), 
while for the submillimeter domain, we use the sensitivity curve that has been estimated for a R=1000 resolution spectrometer at the focal plane of the
CCAT telescope. Details on these instruments are presented in the Appendix. Moreover, we selected a common integration time of 1 hour 
and a total field of view to be covered by our simulated survey of 0.5 deg$^2$. 
For an instrument, such as SAFARI, with a $2\arcmin \times 2 \arcmin$ field of view, this corresponds to 450 hours of integration time,
to be compared to 4.5 hours for a CCAT spectrometer, assuming a field of view of  $20\arcmin \times 20 \arcmin$.

\begin{figure*}[!t]
\includegraphics[width=9truecm]{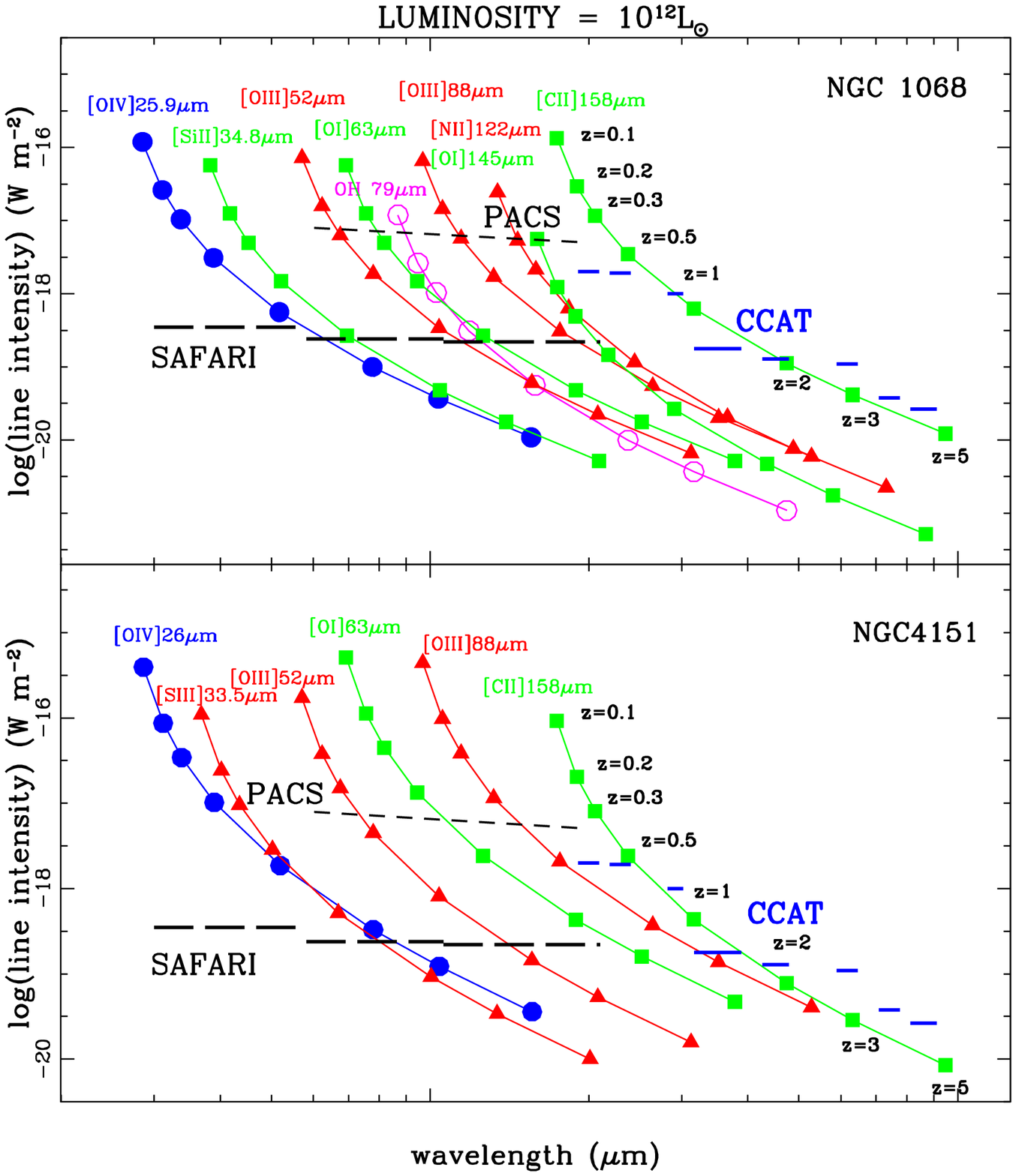} 
\includegraphics[width=9truecm]{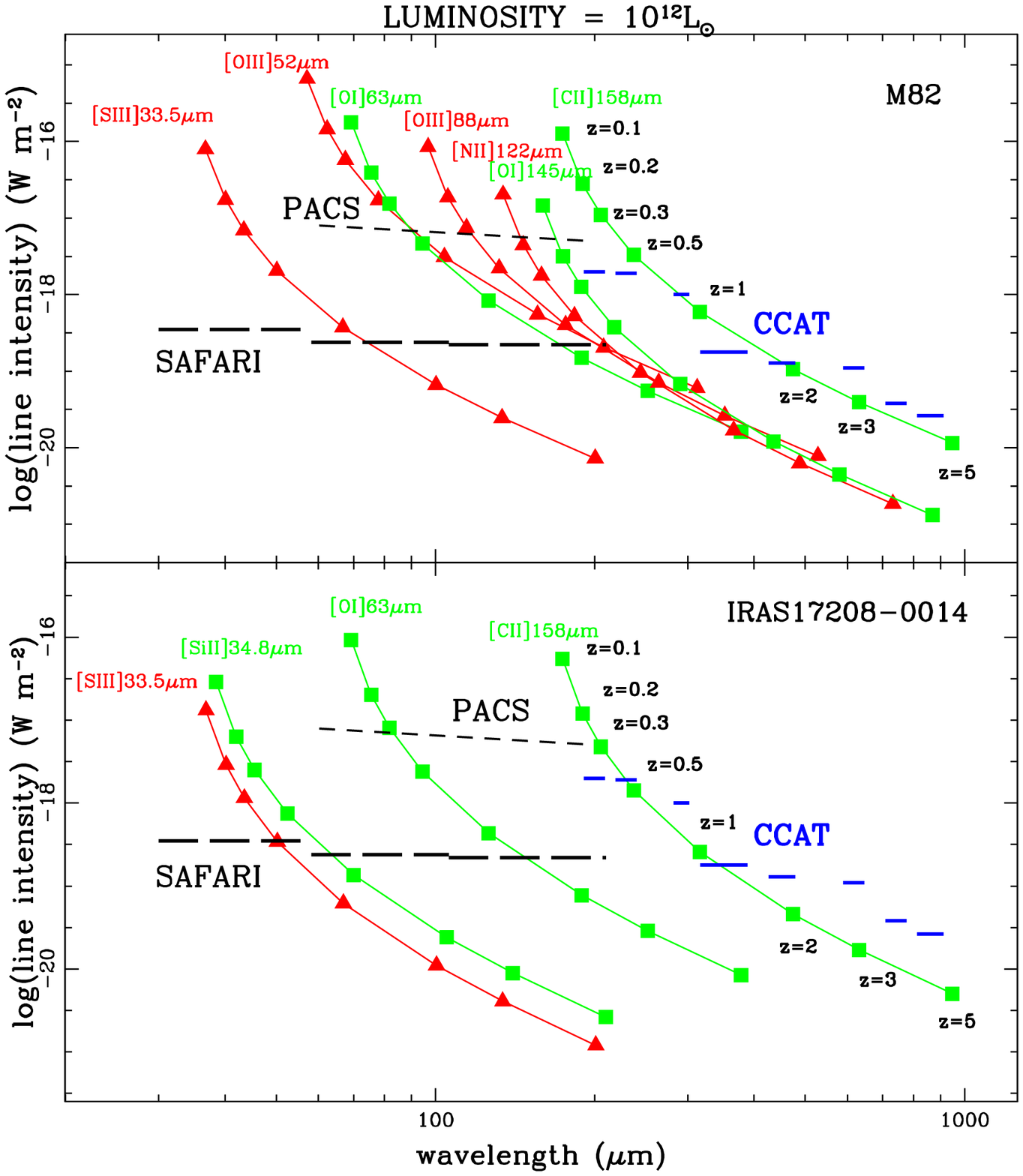} 
\caption{Same as Fig. \ref{fig:template_predictions1}, for long-wavelength lines, where the complementarity between SPICA and CCAT is clearly shown.
For the adopted sensitivities of CCAT we refer to the Appendix. 
}
  \label{fig:template_predictions2}
\end{figure*}

 \begin{figure*}[!t]
   \includegraphics[width=\textwidth]{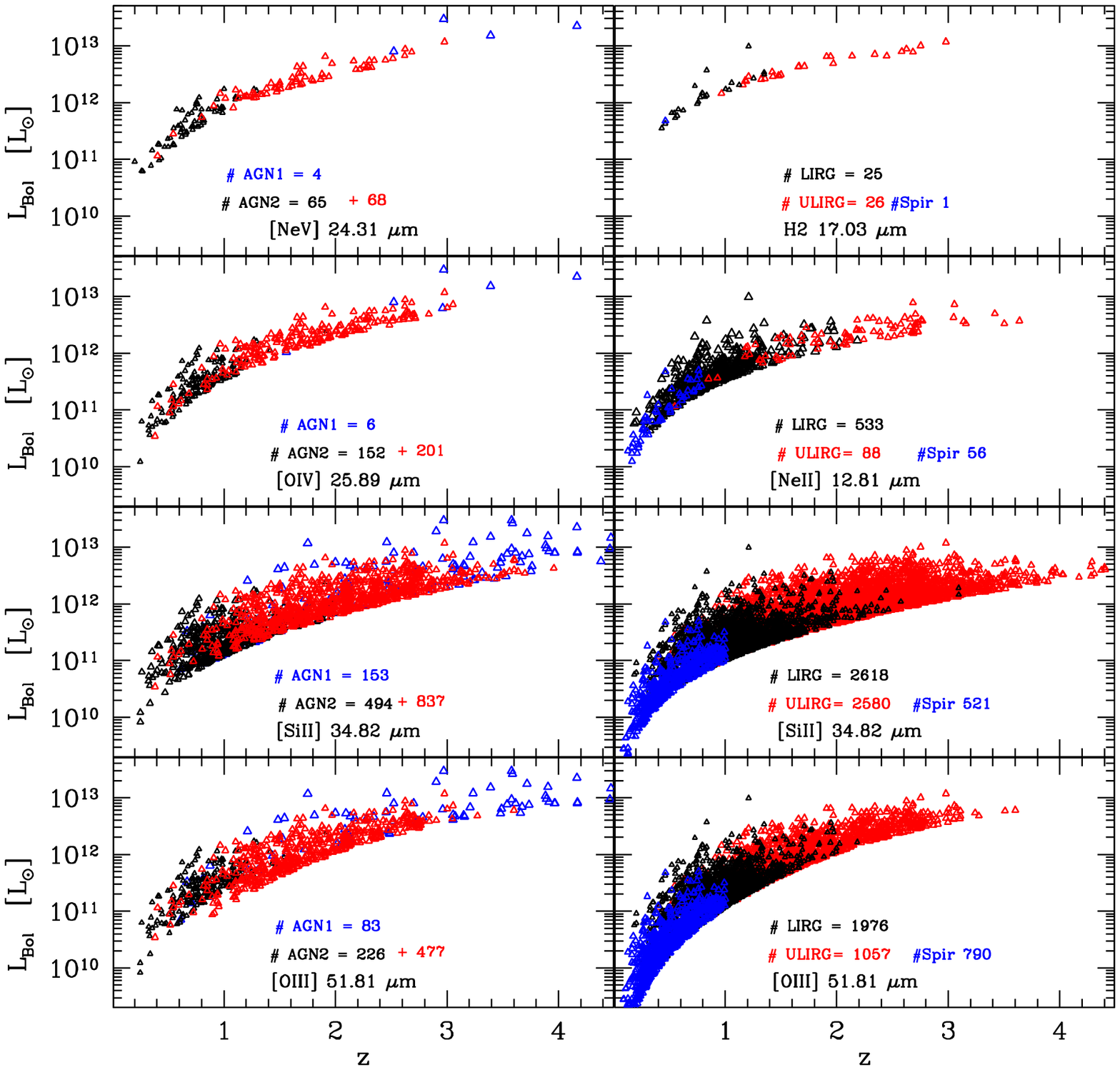} 
    \caption{Prediction of the number of sources of a 0.5 deg$^2$ spectroscopic survey with SAFARI based on \citet{fra10}, giving the number of detectable 
    starburst galaxies (divided by \lir ) and AGN (divided by obscuration) at the 3$\sigma$ level. The adopted line flux sensitivities as a function of wavelength are given in the Appendix.
    The left panels correspond to the AGN predictions, the right panels to the starburst predictions. As for the former, the number of type 2 AGNs associated with the LIRG and the ULIRG populations are shown separately (in black and red, respectively).
    }
 \label{fig:ncounts1}
 \end{figure*}

The resulting number of AGN and starbursts that will be detectable in each line as a function of $z$ with SAFARI is presented in 
Tables~\ref{tab:frances_agn} and~\ref{tab:frances_sb}, respectively, for the \citet{fra10} model. The same results are presented in 
Tables~\ref{tab:gruppioni_agn} and~\ref{tab:gruppioni_sb} for the \citet{gru11} model. The number of detectable galaxies based 
on the \citet{val09} model is presented in Table~\ref{tab:valiante}. 
We note that for the sake of completeness, and to assist further planning and designing of new instrumentation, we also 
present in Tables 1-5 the predicted number of sources that are detectable in the low$-$Z bins for the short-wavelength lines and those 
in the high-$Z$ bins for the long-wavelength lines, even lines outside the nominal SAFARI spectral range. We have simply assumed 
a flat extrapolation of the SAFARI sensitivities to shorter and longer wavelengths. 
Tables~\ref{tab:frances_ccat},~\ref{tab:gruppioni_ccat}, and~\ref{tab:valiante_ccat} present the simulation results for CCAT. 

A basic result of this analysis is that the total number of detectable objects agrees to within a factor of 2$-$3 for most lines and $z$ ranges, 
and that at least a thousand galaxies will be simultaneously detected in four lines at 5$\sigma$ over a half square degree.
A comparison of the output of the three models is plotted in Fig.~\ref{fig:ncounts_hist}  for SAFARI. A survey of the assumed sensitivity
will detect bright lines (e.g., \oi\ and \oiii ) and PAH features in thousands of galaxies at $z$$>$1. Hundreds of $z$$>$1
AGN will be detected in the \oiv\ line, and several tens of $z$$>$1 sources will be detected in \nev\ and \htwo . For \htwo\ in particular, 
this number corresponds to a lower limit. Our models do not account for an increase of the \htwo\ emission efficiency as a cooling mechanism, 
or for the \htwo\ mass content to increase with increasing $z$ and decreasing metallicity. Neither do the applied continuum-to-line luminosity relations 
include sources of extremely high L(\htwo )/\lir\ ratios, associated with shock fronts due to galaxy collisions or AGN feedback mechanisms 
\citep{cluver10,ogle10}. 

Our line detectability results are sensible when compared with predictions from local galaxy templates. To make this comparison, we used four 
objects with well-determined MIR and FIR spectra. These are NGC1068, a prototypical Seyfert 2 AGN, NGC4151, a well-studied type 1 AGN, 
the prototypical moderate-luminosity starburst M82, and a starburst-dominated ULIRG, IRAS17208-0014. Their line intensities are taken from 
\citet{ale, s05}, \citet{stu99, spi97}, \citet{far07, bra08} and \citet{fo01, co99}, respectively for the four templates. 
We scaled the bolometric IR luminosity of all systems to 10$^{12}$ \lsun, and show which lines can be observed as a function of $z$ in Figs.~\ref{fig:template_predictions1} and \ref{fig:template_predictions2}. This basic comparison confirms that faint lines can be detected 
in $z$$\sim$1 ULIRGs with SAFARI. Emission from [\ion{O}{4}] will be observable in z$\sim$3 ULIRGs with SAFARI, while CCAT will observe \cii\ 
out to z$\sim$5.

\begin{figure}[!t]
   \includegraphics[width=\columnwidth]{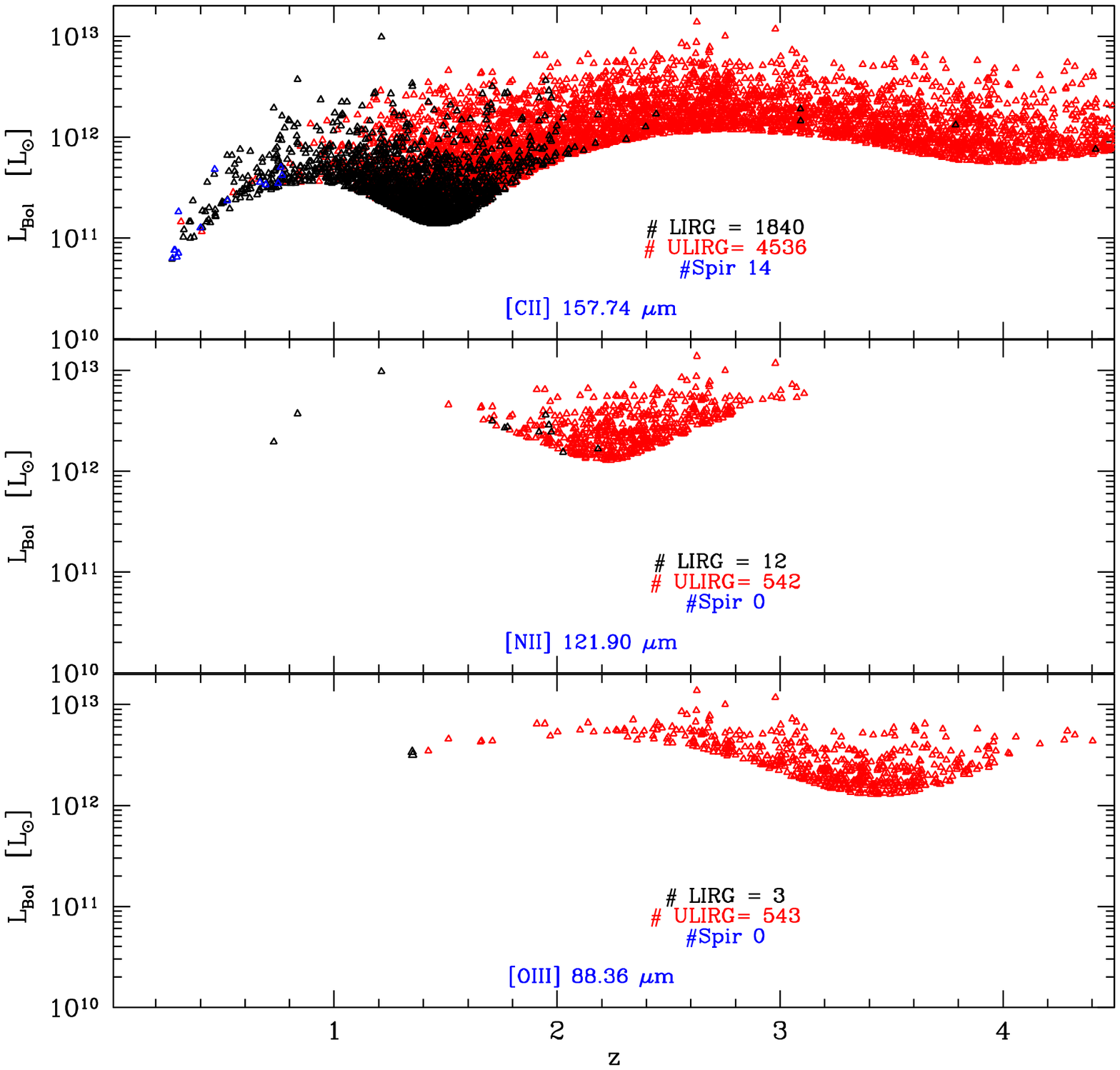} 
    \caption{Prediction of the number of sources of a 0.5 deg$^2$ spectroscopic survey based on \citet{fra10}, giving the number of detectable 
    galaxies at the 3$\sigma$ level with CCAT. The adopted line flux sensitivities as a function of wavelength are reported in the Appendix.
    }
 \label{fig:ncounts2}
 \end{figure}

In Fig.~\ref{fig:ncounts1}, and~\ref{fig:ncounts2} we further investigate in what type of IR-bright galaxies can each line be detected as a function of $z$. We only make this 
comparison for the \citet{fra10} model, as the three models do not differ in terms of their IR-luminosity classification, and as the total number of sources 
is comparable in all cases. We find that for the brightest lines, such as \sitwo , SAFARI will be able to observe LIRGS even at $z$$>$3. However,
for the typical \neii\ and \oiii\ 52$\um$ line luminosities, the transition from LIRGs to ULIRGs will occur at $z$$\sim$2, and for the \nev\ and \htwo\ S1 lines 
at $z$$\sim$1. CCAT will be highly complementary to SPICA, as it will be able to observe the \oiii\ 88$\um$ line at $z$$>$1.3, where this line leaves
the spectral range of SAFARI. We also find that CCAT will be a most efficient instrument for studies of \cii , an important coolant of the interstellar
medium, at all $z$$<$5. At  3$<$z$<$4 alone, it will detect more than 300 galaxies at $5$$\sigma$ level in a 0.5 deg$^2$ survey.

 \begin{figure}
   \includegraphics[width=\columnwidth]{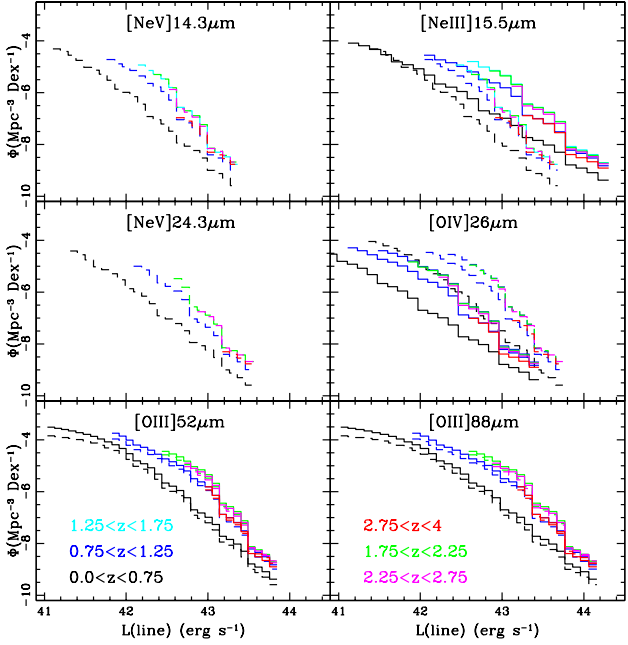}
    \caption{Predicted line luminosity functions of [NeV]14.3$\mu$m, [NeIII]15.5$\mu$m, [NeV]24.3$\mu$m, [OIV]26$\mu$m, [OIII]52$\mu$m and [OIII]88$\mu$m, 
    for SAFARI. Dashed lines correspond to the predictions for AGN, while solid lines correspond to the predictions for starbursts. Where available, the comparison with the observed local LF of AGNs from \citet{tom10} is given. 
 }
 \label{fig:lf1}
 \end{figure}

 \begin{figure}
   \includegraphics[width=\columnwidth]{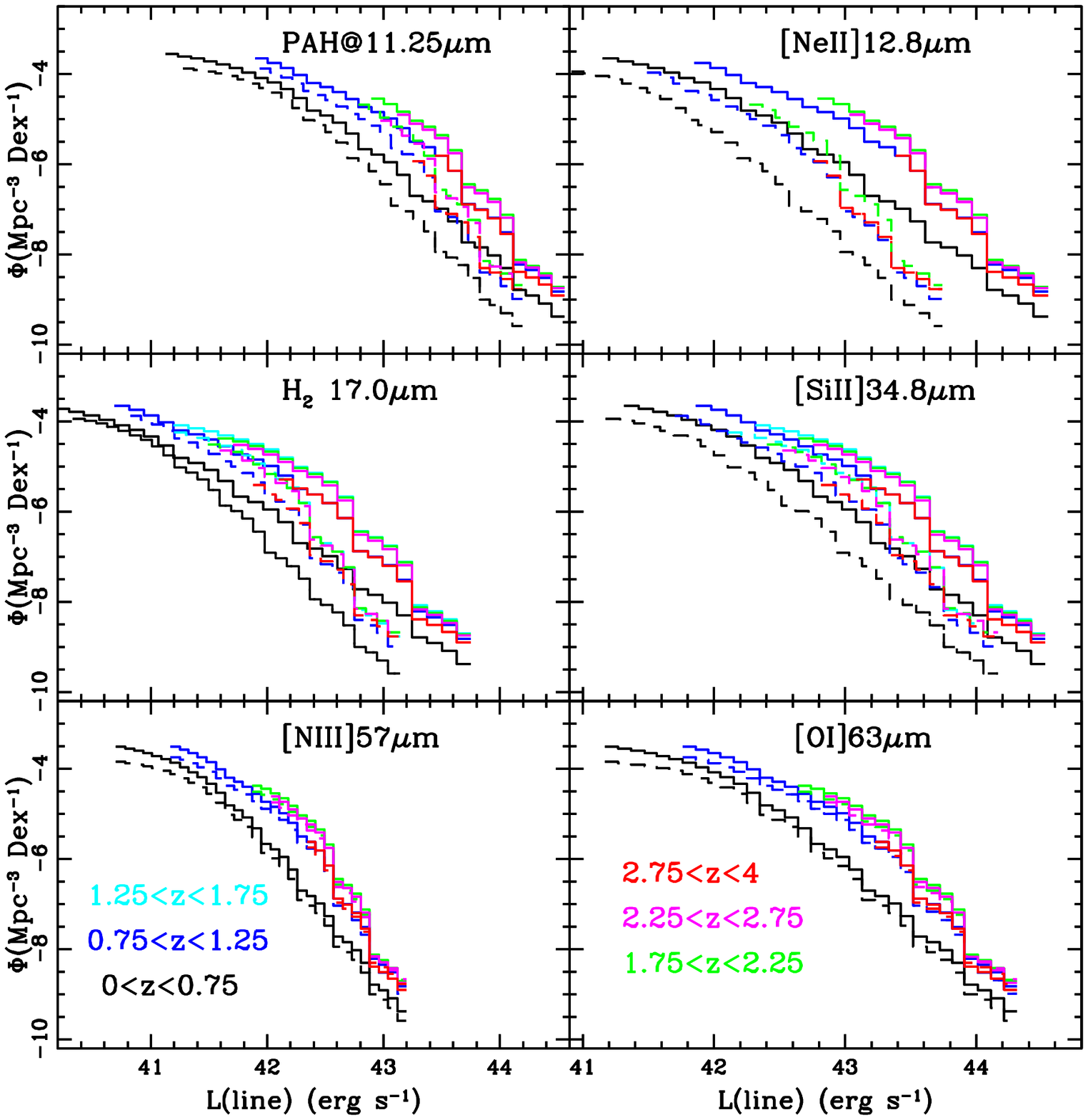}
    \caption{Predicted line luminosity functions of PAH 11.25$\mu$m, [NeII]12.8$\mu$m, H$_2$ 17$\mu$m, [SiII]34.8$\mu$m , [NIII]57$\mu$m and [OI]63$\mu$m, 
       for SAFARI. Dashed lines correspond to the predictions for AGN, while solid lines correspond to the predictions for starbursts. Where available, the comparison with the observed local LF of AGNs from \citet{tom10} is given. 
 }
 \label{fig:lf2}
 \end{figure}

\subsection{Line luminosity function predictions in the IR/submm}

We present in Fig.~\ref{fig:lf1} and ~\ref{fig:lf2}
the predicted luminosity functions of AGN and starburst galaxies 
for each line and feature in the \cite{fra10} model.
It is clear from the figure that the AGN line luminosity functions (dashed lines in the figure) for the lowest redshift range (0.$<$z$<$0.75) are 
in agreement with the local luminosity functions, at an average $<z>$ of 0.03.
The space densities of starburst galaxies are expected to be higher than those of AGN, for any line, except for the [OIV] line,
which--as is well known--is much fainter in starburst galaxies.
The volume densities of AGN drop faster with $z$ than 
those of starbursts when traced by the \neii\ and \neiii\ lines, possibly due to a saturation of the lines in AGN of high luminosities. The total 
number of AGN detected in the same lines is one to two orders of magnitude lower than that detected for starbursts at any $z$. The number 
of sources detected in \oiii\ 88$\um$ is comparable for AGN and starbursts at all $z$. This result most likely differs from that for \neiii\, which
comes from an ion of comparable ionization potential to \ion{O}{3}, because of the single relationship used to convert the line luminosity to \lir\
for long wavelength lines. This is even true for the \oi\ 63$\um$ line, which also has a high critical density for collisional de-excitation, 
$\sim$10$^{6}$ cm$^{-3}$, making it bright in AGN. The contamination of the \oiv\ AGN luminosity functions from starbursts is minimal all the 
way through $z$=4. 
The volume density of actively accreting black holes as traced by \oiv\ and \nev\  increases up to 
1$\lesssim$$z$$\lesssim$3 before it drops back down at $z$=4. The same applies for tracers of star-formation, reproducing the suggested 
coevolution of black hole growth and stellar mass build-up.

\section{Discussion: new parameter-space coverage by SPICA and CCAT}
\label{sec:discussion}

Which of the questions raised in the Introduction will the proposed future telescopes address? The lines predominantly emitted
by ions in star-forming complexes, like \sitwo , \cii , and \neii , will be detected in \lir $>$10$^{11}$\lsun\ systems at least out to $z$$\sim$2. 
They will also be detected in 10$^{10}$\lsun $<$\lir $<$10$^{11}$\lsun\ galaxies at least out to $z$$\sim$1. This result indicates that 
star-formation tracers will be detected and compared for the sources that are mainly responsible for the formation of present-day 
ellipticals. The creation of line-ratio diagnostic diagrams, and the comparison of the line to IR continuum luminosities in hundreds 
of sources will help us, in combination with imaging data, to further address the star-formation bimodality, and to obtain a more coherent picture 
of the intermediate/high-$z$ IR-bright populations. Several sources will be detected with CCAT even past the peak of star-formation activity 
at $z$$>$3 in the strong \cii\ line, which will help us constrain the shape of Lilly-Madau diagrams at such high redshifts.

To date, the detection of FIR fine-structure lines has only been achieved in lensed $z$$>$1 systems. \textit{Herschel} PACS detected the \oiii\ 52$\mu$m 
line with a flux of 9$\times$ $10^{-19}  {\rm W m^{-2}}$ in IRAS~F10214+4724 at \textit{z}=2.28 \citep{stu10}.  In a $z$=1.32 source, MIPS 
J142824.0+352619, \oiii\ 52$\mu$m and \oi\ 63$\mu$m were detected with fluxes of 3.7 and 7.8$\times$ $10^{-18} {\rm W m^{-2}}$, respectively  
\citep{stu10}. The \oiii\ 88$\mu$m line was detected at the Caltech Submillimeter Observatory (CSO) with the ZEUS spectrometer \citep{ferk10b}
in APM~08279+5255 at $z$=3.9 and SMM~J02399-0136 at $z$=2.8 \citep{ferk10}, with fluxes 
of 2.68 and 6.04 $\times$$10^{-18} {\rm W m^{-2}}$. All four systems are lensed galaxies, with magnification factors estimated to be
in the range 2.4$-$90 \citep{ega00,ao08,rie09,ivi10b}.  Such experiments were not possible for unlensed galaxies with the present-day 
missions, leaving ample room for new discoveries for SPICA and CCAT.

The recent detection of [NII]122$\mu$m \citep{ferk11} and [CII]158$\mu$m \citep{sta10} lines in a few high redshift galaxies with the ZEUS 
spectrometer \citep{ferk10b} at the CSO shows that these lines can be much brighter than in local galaxies. The [NII]122$\mu$m line to FIR
luminosity ratio is 2-10 times higher in the two observed galaxies (H1413+117 and SMMJ02399-0136) than in the local galaxies that we 
used for deriving the line to continuum relations adopted for our predictions. Similarly the [CII]/FIR luminosity ratio in the 1 $<$ $z$ $<$ 2
galaxies can be twice as high as the value observed in the local Universe, despite having a large rage from 0.024\% to 0.65\% . This indicates
that our estimates can be considered conservative, and that CCAT could observe many more galaxies than what shown in our predictions. It
could potentially trace the star formation history of galaxies back to very high redshift.

A simple conversion of the SAFARI sensitivity limit into an \htwo\ line flux and mass indicates that detection of \htwo -bright galaxies at $z$$>$6 
will be feasible over large areas. An S1 flux of 2.5$\times$10$^{-19}$ W m$^{-2}$ at a typical temperature of 300K, yields an \htwo\ mass of 
3$\times$10$^{10}$\msun\ at $z$=6. Cosmological simulations indicate that $z$$\sim$6 galaxies of such a molecular gas mass can exist \citep{obr09}.
Unless there is a significant increase in the efficiency of \htwo\ to cool the ISM, or unless there is a considerable number of sources with an
AGN jet-ISM interaction that leads to very high L(\htwo )/\lir\ ratios \citep{ogle10}, our conclusion will not hold at $z$$\sim$10. At that $z$, the same 
temperature and S1 flux correspond to a mass of 10$^{11}$\msun . Even though the \htwo\ (0-0)S0 line at 28.03$\um$ will be in the spectral range 
of CCAT at $z$$>$ 6, its detection will be equally hard: assuming an S0 flux of 2$\times$10$^{-19}$ W m$^{-2}$ and an ortho-to-para ratio of 3, the 
minimum detectable \htwo\ mass would be $\sim$10$^{11}$\msun\ at $z$=6. SPICA will thus be unique for \htwo\ studies seeking the end of the reionization era.

Another parameter space unique to SPICA will be the detection of resolved \oiv\ 25.89$\um$ emission up to $z$$\sim$4. The calibration of the widths 
of high-ionization MIR lines, as probes of the narrow-line-region kinematics, to the black hole mass was attempted with \spi\ for sources out to $z$=0.3 
\citep{Dasyra2008,Dasyra2011}. It might be useful for the mid-infrared instrument of JWST, which will be able to observe the \nev\ 14.32$\um$ line out to $z$$\sim$0.8. 
The breakthrough for such studies will come however from SPICA. SAFARI will access the $z$$>$1 Universe and go past its accretion-rate-history
peak. The shape of the accretion rate functions will be then compared to the shape of the mass functions of the obscured and unobscured black holes upon which 
the material is accreted. The inclusion of obscured black holes in mass functions of unobscured AGN \citep[e.g.,][]{vest09} could change their observed volume-density 
normalization factors and shapes. The most obscured AGN are thought to be missing from a redshift of only 1 \citep[e.g.,][]{gilli07},
unlike the bulk of the black hole growth that is thought to be occurred at $z$ $>$ 1.

\section{Summary}

We used three galaxy formation and evolution models, constrained by luminosity functions from multiwavelength observations, to  predict the 
number of  star-forming and active galaxies that are observable in any IR luminosity bin with look-back time. We converted the number counts of 
galaxies per IR luminosity to number counts of galaxies per line luminosity using several (new) line-to-continuum conversion relations, built upon 
local AGN and starburst galaxy samples. We compute our results for an hour-long integration/FoV half-square-degree survey, and for the sensitivity values of SPICA
SAFARI and CCAT. Their anticipated values are 2.5$\times$10$^{-19}$ W m$^{-2}$ at 160$\um$ and 1.1$\times$10$^{-19}$ W m$^{-2}$ at 620$\um$, 
respectively. These telescopes/instruments were selected because they are designed to be survey machines, able to perform large blind cosmological surveys
in reasonable integration times. We find that SAFARI will detect thousands of $z$$>$1 galaxies in bright low-ionization fine-structure lines such as 
\sitwo\ and \neii , and several tens of $1<z<2$ galaxies in fainter lines such as \oiv, \nev\ and \htwo\ (0-0)S1. For the bright lines, normal galaxies will be 
observed out to $z$$\sim$1, LIRGs out to $z$$\sim$2, and ULIRGs to even higher $z$. This means that  studies of the ionized gas properties in the galaxies 
that form the present day massive ellipticals will be feasible. AGN/star-formation diagnostic diagrams will be obtained for different classes of IR-bright 
galaxies at $1<z<2$, which will enable us to not only look for a redshift evolution of line ratios, but also for a luminosity evolution within each $z$ range. 
Further tests of the black hole growth - galaxy build-up coevolution scenario will be performed, as the creation of accretion-rate functions and mass 
functions will be determined out to $z$$\sim$4, for both obscured and unobscured black holes using the \oiv\ line. Over large areas, SAFARI could also be 
able to detect \htwo\ at $z$$\gtrsim$6, and help constrain the end of the reionization era. 
In the light of the current findings at ground-based submillimeter telescopes 
of substantially brighter fine structure lines in high redshift galaxies compared to local galaxies, 
we are even more confident that CCAT will be unique for studying the star formation history of galaxies
back to very high redshift.

\acknowledgments

We acknowledge input from Takao Nakagawa, PI of the SPICA Mission, and from Peter Roelfsema, 
Frank Helmich and Bruce Swinyard, PI and members of the SAFARI Consortium, respectively. We also
acknowledge Gordon Stacey, Simon Radford, Jason Glenn and Riccardo Giovanelli for information
on the CCAT project and its planned instrumentation. We thank Matt Malkan who commented on this manuscript  
and Scott Douglas, Nicola Sacchi, Silvia Tommasin, Anna Di Giorgio, John Scige Liu and Erina Pizzi for assisting us 
in improving this document.  We also thank the anonymous referee for a very thorough and constructive report. 
This work is based on observations made with the \spi\ Space Telescope which is operated by the Jet Propulsion 
Laboratory and Caltech under a contract with NASA.  K. M. D. acknowledges support by the European Community 
through a Marie Curie Fellowship (PIEF-GA-2009-235038)  awarded under the Seventh Framework Programme (FP7/2007-2013).


\begin{appendix}
\section{Description of the SPICA mission and the CCAT facility to be used for future, blind cosmological surveys in the IR and submm}
\label{sec:appendix}

The deep cosmological surveys undertaken by ISO \citep{kes96}, \spi\ \citep{wer04}, AKARI \citep{mur07,goto10}, WISE \citep{wri10}, and {\it Herschel} 
\citep{pil10} will have produced catalogues containing the fluxes of many tens of thousands of IR-bright sources by the early 2020's. These 
catalogues will provide excellent targets to be followed up by ALMA \citep{bro04, woo08}, JWST \citep{gar06}, SPICA \citep{sw09}, and 
CCAT \citep{seb10}. Among the listed facilities, JWST and ALMA will be more suited to deep follow-up spectroscopy of known targets. SPICA and 
CCAT will be suited for performing blind large-scale spectro-photometric surveys, because of the wide field of view (of several arcminutes squared) 
of their instruments. Covering different wavelength ranges, these two instruments will be highly complementary. In the rest of the Appendix, 
we provide details of their present-day design concepts.


SPICA is a proposed JAXA-led astronomical mission with suggested contributions by European, Korean, and possibly
US institutions, and with a launch date planned in the early 2020s \citep{sw09}. With a 3-m mirror that is actively cooled to $<$6K,
and a state-of-the art focal plane instrument suite, SPICA will make imaging and spectroscopic observations 
over the 5-210$\mu$m spectral range with unprecedented sensitivity. It will offer an improvement in raw photometric sensitivity 
with respect to \textit{Herschel}  of two orders of magnitude in the FIR. 
In the MIR, it will extend the capabilities of JWST with an uninterrupted spectral coverage in the range 5-38$\mu$m.
The Mid Infrared Camera and Spectrometer (MCS) planned for SPICA \citep{wa10,spica} is an 
integral field unit with a field of view of 12$\arcsec$ $\times$ 6$\arcsec$ at 10-20$\mu$m and 12$\arcsec$ $\times$ 
12.5$\arcsec$ at 19.5-36.1$\mu$m. Its 5 $\sigma$, 1 hour sensitivity is in the range 2-2.5 $\times 10^{-20} W m^{-2}$ 
\citep{wa10}. Observing capability in the FIR is provided for SPICA by SAFARI \citep{sw09,spica}. 
Proposed by a consortium of European institutes (with Canadian and Japanese participation), SAFARI is an imaging 
Fourier Transform Spectrometer (FTS) with a field of view of 2$\arcmin$ $\times$ 2$\arcmin$.  The FTS provides an instantaneous 
spectral coverage of the 34 - 210$\mu$m wavelength range and spectral resolution modes with $\lambda$/$\Delta \lambda$
of 2000 (at 100 $\mu$m), $\sim$ few hundred, or even as low as 20 $<$ $\lambda$/$\Delta \lambda$ $<$ 50. With its sensitive 
superconducting transition edge sensor (TES) detectors \citep{kho10}, SAFARI will offer a factor of $\sim$100 increase in raw sensitivity 
in the continuum, and $\sim$15 in high-resolution mode spectroscopy compared to \textit{Herschel} PACS \citep{pog10}. The 
improvement in spectral mapping speed over PACS will be of more than 100 at $\lambda$/$\Delta \lambda$$\sim$ 2000. Assuming 
a  Noise Equivalent Power 
(NEP) of $2 \times 10^{-19} W/\sqrt{Hz}$ for the SAFARI detectors, the 5 $\sigma$, 1 hour detection limits are predicted 
to be 4.16,  2.58, 1.89 and 2.48 $\times 10^{-19} {\rm W m}^{-2}$ for the four planned spectral bands centred at 48, 85, 135 and 160$\mu$m, 
respectively, for an unresolved line at $\lambda$/$\Delta \lambda$$\sim$ 2000  (Swinyard, B., 2011, priv. comm.).
 
The Cerro Chajnantor Atacama Telescope (CCAT) will be a 25m submm wave telescope to be constructed near the summit of Cerro 
Chajnantor \citep{seb10}. The CCAT science case, observatory requirements, and conceptual design were developed as part of 
a study jointly funded by Cornell and Caltech/JPL, which resulted in a Feasibility/Concept Design Study Report, available at: 
{\it http://www.submm.org/doc/2006-01-ccat-feasibility.pdf}. The design of the CCAT telescope is currently undergoing changes to
increase the field of view from 20$\times$20 arcminutes square to 1 square degree \citep{seb10}. This would substantially increase 
the survey capability of the telescope. CCAT will carry out spectroscopic surveys of submm galaxies, using multi-object versions of 
broadband direct-detection grating spectrometers such as Z-Spec \citep{bra04} and ZEUS \citep{ferk10b}, now in use at the CSO. 
Conceptual development indicates that spectrometers capable of observing 10-100 objects simultaneously while spanning multiple 
atmospheric windows will be feasible \citep{sta06} . 
Sensitivity estimates based on the 25 m telescope CCAT telescope with a 10 $\mu$m rms surface 
on Cerro Chajnantor (5600 m elevation) for a spectrometer with a resolution of 1000 (5$\sigma$, 1 hour) that include wavelength-dependent,
typical precipitable water vapor corrections are: 
2.0, 1.9 and 1.0 $\times 10^{-18} {\rm W m}^{-2}$ at 200$\mu$m, 230$\mu$m and 291$\mu$m respectively,
1.8, 1.3 and 1.1 $\times 10^{-19} {\rm W m}^{-2}$ at 350$\mu$m, 450$\mu$m and 620$\mu$m respectively and
3.8 and 2.6 $\times 10^{-20} {\rm W m}^{-2}$ at 740$\mu$m and 865$\mu$m respectively (Stacey, G. 2011, private comm.).



\end{appendix}


\clearpage

\begin{deluxetable}{lccccccc}
\tabletypesize{\scriptsize}
\tablewidth{0pt} 
\tablecaption{Number of AGN detectable in a SAFARI survey of 0.5$^2$ 
in IR lines as a function of redshift at 5$\sigma$ (3$\sigma$) in 1 hr. integration per FoV, following \citet{fra10} \label{tab:frances_agn} } 
\tablehead{ \colhead{line/redshift}& \colhead{0$<$z$<$0.75}  & \colhead{0.75$<$z$<$1.25}   & \colhead{1.25$<$z$<$1.75} & \colhead{1.75$<$z$<$2.25 }  & \colhead{2.25$<$z$<$2.75}  & \colhead{2.75$<$z$<$4}  &  \colhead{all redshifts \#} } 
\startdata
PAH(11.25$\mu$m)               & 687. (945.) \dag & 633. (1449.) \dag & 439. (703) \dag & 54.0 (138.)    &  5.85 (47.7) & \nodata (1.35)       & 60.0 (187.) \ddag \\ 
$\left[ NeII \right]$ 12.81$\mu$m  & 152. (366.) \dag & 122. (247.) \dag & 14.4 (45.0)  & 0.90 (6.75)  &  \nodata (0.45) &  \nodata (\nodata)  & 15.3 (52.2) \ddag \\
$\left[ NeV \right]$  14.32 $\mu$m & 43.2 (152.) \dag & 20.2 (82.3) \dag & 0.45 (5.85) & \nodata (\nodata)&\nodata (\nodata)&\nodata (\nodata) & 0.45 (5.85) \ddag \\
$\left[ NeIII \right]$  15.55$\mu$m & 152. (366.) \dag & 35.1 (82.3)     & 5.85 (45.0)  & 0.45 (1.80) &  \nodata (0.45) &  \nodata (\nodata)   & 41.4 (130.) \ddag \\
H$_2$(17.03$\mu$m)             & 10.8 (43.2) \dag & \nodata (6.75)& \nodata (0.45) & \nodata (\nodata) & \nodata (\nodata) & \nodata (\nodata) & \nodata (7.20) \ddag \\
$\left[ SIII \right]$  18.71$\mu$m & 43.2 (106.) \dag & 3.15 (11.2) & \nodata (0.90) & \nodata (\nodata) &\nodata (\nodata) &\nodata (\nodata) & 46.3 (118.)\ddag \\
$\left[ NeV \right]$  24.32$\mu$m   & 26.6 (69.8)      & 11.2 (55.8)     & 14.4 (45.0)  & 0.90 (6.75)  & \nodata (0.90) &  \nodata (\nodata)   & 53.1 (178.)\\
$\left[ OIV \right]$  25.89$\mu$m   & 152. (366.)      & 82.3 (176.)     & 73.8 (246.)  & 17.1 (54.0)    &  0.90 (14.8) &  \nodata (\nodata)   & 326. (857.)\\
$\left[ SIII \right]$  33.48$\mu$m  & 69.8 (210.)      & 122. (247.)     & 45.0 (174.)  & 6.75 (32.8)    &  0.90 (14.8) &  \nodata (\nodata)   & 244. (679.)\\
$\left[ SiII \right]$  34.81$\mu$m  & 210. (366.)      & 333. (633.)     & 174. (439.)  & 54.0 (204.)    &  14.8 (121.) &  1.35 (11.7)         & 787. (1775.)\\
$\left[ OIII \right]$  51.81$\mu$m  & 464. (687.)      & 333. (633.)     & 246. (563.)  & 54.0 (204.)    &  14.8 (28.8) & \nodata (0.90) \dag  & 892. (2117.) \ddag \\
$\left[ NIII \right]$  57.32$\mu$m  & 106. (281.)      & 55.8 (176.) & 14.4 (73.8)  & \nodata (0.90) &  \nodata (\nodata) &  \nodata (\nodata) & 176. (532.)\\
$\left[ OI \right]$  63.18$\mu$m    & 687. (945.)      & 1128. (1817.)   & 563. (863.)  & 204. (381.)    &  47.7 (179.) &  4.50 (22.5) \dag    & 2833. (4184.) \ddag \\
$\left[ OIII \right]$  88.35$\mu$m  & 687. (945.)      & 459. (856.)  & 246. (563.)  & 88.2 (204.) \dag &  28.8 (77.4) \dag&  0.90 (11.7) \dag & 1392. (2364.) \ddag \\
\enddata
\tablenotetext{*}{Notes: \#: total number of AGN inside detectable the SAFARI spectral range; 
\dag: outside the SAFARI spectral range;\ddag: excluding detections outside the SAFARI spectral range.}
\end{deluxetable}

\begin{deluxetable}{lccccccc}
\tabletypesize{\scriptsize}
\tablewidth{0pt}
\tablecaption{Number of starburst galaxies detectable in a SAFARI survey of 0.5$^2$  
in IR lines as a function of redshift at 5$\sigma$ (3$\sigma$) in 1 hr. integration per FoV, following \citet{fra10} \label{tab:frances_sb} } 
\tablehead{ \colhead{line/redshift}& \colhead{0$<$z$<$0.75}  & \colhead{0.75$<$z$<$1.25}   & \colhead{1.25$<$z$<$1.75} & \colhead{1.75$<$z$<$2.25 }  & \colhead{2.25$<$z$<$2.75}  & \colhead{2.75$<$z$<$4}  &  \colhead{all redhifts \#} } 
\startdata
PAH(11.25$\mu$m)                 & 1023. (1519.) \dag&974. (1790.) \dag& 608. (984.) \dag & 119. (277.) &  39.1 (105.) &  1.80 (15.7)      & 160. (398.) \ddag \\ 
$\left[ NeII \right]$ 12.81$\mu$m & 636. (1023.) \dag&692. (1340.) \dag& 239. (460.)      & 72.9 (187.) &  20.2 (64.3) &  0.90 (6.30)      & 333. (718.) \ddag \\ 
$\left[ NeIII \right]$  15.55$\mu$m & 115. (254.) \dag  & 80.5 (118.)  & 37.3 (101.)      & 9.00 (44.1) &  1.35 (8.10) &  0.45 (1.80)      & 129. (273.) \ddag \\
H$_2$(17.03$\mu$m)                 & 28.8 (43.2) \dag  & 9.90 (29.7)  & 2.25 (7.65) & 0.45 (1.80) & \nodata (\nodata) & \nodata (\nodata)  & 12.6 (39.1)\ddag \\
$\left[ SIII \right]$  18.71$\mu$m  & 176. (355.) \dag  & 50.8 (118.)  & 19.3 (61.6)      & 2.70 (22.9) &  1.35 (8.10) &  \nodata (0.45)   & 250. (566.)\ddag \\
$\left[ OIV \right]$  25.89$\mu$m   & 2.70 (9.45)       & 1.80 (9.90)  & 1.35 (7.65) & 0.45 (0.90) & \nodata (\nodata) & \nodata (\nodata) & 6.30 (27.9)\\
$\left[ SIII \right]$  33.48$\mu$m  & 355. (636.)       & 495. (974.)  & 337. (608.)      & 119. (277.) &  64.3 (164.) &  6.30 (30.1)      & 1377. (2689.)\\
$\left[ SiII \right]$  34.81$\mu$m  & 482. (816.)       & 974. (1790.) & 608. (984.)      & 277. (519.) &  164. (339.) &  30.1 (81.4)      & 2535. (4529.)\\
$\left[ OIII \right]$  51.81$\mu$m  & 816. (1257.)      & 495. (974.)  & 337. (783.)    & 72.9 (277.) &  2.25 (39.1) & \nodata (0.90) \dag & 1723. (3330.) \ddag \\
$\left[ NIII \right]$  57.32$\mu$m  & 176. (482.)  & 80.5 (258.)  & 19.3 (101.) & \nodata (1.80) & \nodata (\nodata) &  \nodata (\nodata)  & 276. (843.)\\
$\left[ OI \right]$  63.18$\mu$m    & 1257. (1808.)    & 1790. (2960.)& 783. (1221.)     & 277. (519.) &  64.3 (242.) &  6.30 (30.1) \dag  & 4171. (6750.) \ddag \\
$\left[ OIII \right]$  88.35$\mu$m  & 1257. (1808.) & 692. (1340.) & 337. (783.) & 119. (388.) \dag & 39.1 (105.) \dag &  0.90 (15.7) \dag & 2286. (3931.) \ddag \\
\enddata
\tablenotetext{*}{Notes: \#: total number of AGN inside detectable the SAFARI spectral range; 
\dag: outside the SAFARI spectral range;\ddag: excluding detections outside the SAFARI spectral range.}
\end{deluxetable}

\begin{deluxetable}{lccccccc}
\tabletypesize{\scriptsize}
\tablewidth{0pt} 
\tablecaption{Number of AGN detectable in a SAFARI  survey of 0.5$^2$  
in IR lines as a function of redshift at 5$\sigma$ (3$\sigma$)  in 1 hr.  integration per FoV, following \citet{gru11} \label{tab:gruppioni_agn} } 
\tablehead{ \colhead{line/redshift}& \colhead{0$<$z$<$0.75}  & \colhead{0.75$<$z$<$1.25}   & \colhead{1.25$<$z$<$1.75} & \colhead{1.75$<$z$<$2.25 }  & \colhead{2.25$<$z$<$2.75}  & \colhead{2.75$<$z$<$4}  &  \colhead{all redhifts \#} } 
\startdata
PAH(11.25$\mu$m)                      &1057.(1451.) \dag& 1076. (2317.) \dag & 486. (1126.)\dag& 64.3 (148.)    &  13.9 (75.6) & 0.90  (7.65)   & 79.1 (231.)\ddag \\
$\left[ NeII \right]$ 12.81$\mu$m     & 398.( 701.) \dag&  111. ( 326.) \dag & 9.90 (28.8 )    & 3.60 (12.6)    &  0.45 (1.80) & \nodata (\nodata)  & 14.0 (43.2)\ddag \\
$\left[ NeV \right]$  14.32 $\mu$m    & 177.( 398.) \dag&  9.90 ( 58.0) \dag & 0.90 (5.85 )    & \nodata (0.90)    &  \nodata (\nodata) & \nodata (\nodata)  & 0.90 (6.75)\ddag \\
$\left[ NeIII \right]$  15.55$\mu$m   & 398.( 701.) \dag&  16.6 ( 58.0)      & 5.85 (28.8 )    & 1.80 (6.30)    &  0.45 (1.80) & \nodata (0.45)   & 24.7 (95.3)\ddag \\
H$_2$(17.03$\mu$m)                    & 82.3( 177.) \dag&  0.45 ( 3.15)  & \nodata (0.90 )  & \nodata (\nodata) &  \nodata (\nodata) & \nodata (\nodata)  & 0.45 (4.05)\ddag \\
$\left[ SIII \right]$  18.71$\mu$m    & 177.( 398.) \dag&  1.80 ( 5.40)      & 0.45 (1.80 )    & \nodata (0.45)    &  \nodata (0.45) & \nodata (\nodata)  & 2.25 (10.3)\ddag \\
$\left[ NeV \right]$  24.32$\mu$m     & 124.( 240.)     &  5.40 ( 32.8)      & 9.90 (28.8 )    & 3.60 (12.6)    &  0.45 (3.60) & \nodata (\nodata)  & 143. (318.)\\
$\left[ OIV \right]$  25.89$\mu$m     & 491.( 701.)     &  58.0 ( 192.)      & 47.7 (187. )    & 22.0 (64.3)    &  3.60 (25.2) & 0.90  (4.05)   & 623. (1174.)\\
$\left[ SIII \right]$  33.48$\mu$m    & 240.( 491.)     &  111. ( 326.)      & 28.8 (121. )    & 12.6 (39.1)    &  6.75 (25.2) & 0.90  (4.05)   & 400. (1006.)\\
$\left[ SiII \right]$  34.81$\mu$m    & 491.( 701.)     &  512. (1076.)      & 121. (486. )    & 64.3 (209.)    &  45.9 (170.) & 7.65  (28.3)  & 1242. (2670.)\\
$\left[ OIII \right]$  51.81$\mu$m    & 814.(1184.)     &  512. (1076.)      & 188. (759. )    & 64.3 (209.)    &  6.75 (45.9) & \nodata (\nodata) & 1585. (3274.) \\
$\left[ NIII \right]$  57.32$\mu$m    & 398.( 701.)     &  32.8 ( 192.)      & 9.90 (47.7 )    & 0.45 (3.60)    &  \nodata (0.45) & \nodata (\nodata)  & 441. (945.)\\
$\left[ OI \right]$  63.18$\mu$m      &1057.(1316.)     & 1862. (2805.)      & 759. (1593.)    & 209. (391.)    &  75.6 (230.) & 15.7  (52.6) \dag  & 3978. (6335.)\ddag\\
$\left[ OIII \right]$  88.35$\mu$m    &1057.(1316.)     &  762. (1443.)    & 188. (759. ) & 100. (209.)\dag&  45.9 (117.) \dag & 4.05  (28.3) \dag  & 2007. (3518.) \ddag\\
\enddata
\tablenotetext{*}{Notes: \#: total number of AGN inside detectable the SAFARI spectral range; 
\dag: outside the SAFARI spectral range;\ddag: excluding detections outside the SAFARI spectral range.}
\end{deluxetable}

\begin{deluxetable}{lccccccc}
\tabletypesize{\scriptsize}
\tablewidth{0pt} 
\tablecaption{Number of Starburst galaxies detectable in a SAFARI  survey of 0.5$^2$  
in IR lines as a function of redshift at 5$\sigma$ (3$\sigma$) in 1 hr. integration per FoV, following \citet{gru11}  \label{tab:gruppioni_sb} } 
\tablehead{ \colhead{line/redshift}& \colhead{0$<$z$<$0.75}  & \colhead{0.75$<$z$<$1.25}   & \colhead{1.25$<$z$<$1.75} & \colhead{1.75$<$z$<$2.25 }  & \colhead{2.25$<$z$<$2.75}  & \colhead{2.75$<$z$<$4}  &  \colhead{all redhifts \#} } 
\startdata
PAH(11.25$\mu$m)                 & 515. (779.) \dag& 245. (329.) \dag & 270. (373.) \dag & 71.1 (140.) &  19.8   (48.2)  & 2.70 (10.4)     & 93.6 (199.) \ddag \\
$\left[ NeII \right]$ 12.81$\mu$m    & 326. (515.) \dag& 212. (284.) \dag & 128. (219.)  & 48.1 (102.) &  11.7   (31.5)  & 1.35 (5.85)     & 189. (358.)  \ddag \\
$\left[ NeIII \right]$ 15.55$\mu$m   & 62.1 (115.) \dag& 52.2 (72.0)      & 29.2 (64.8)  & 13.0 (32.4) &  2.25   (7.20)  & 0.45 (2.70)     & 97.1 (179.)  \ddag \\
H$_2$(17.03$\mu$m)                   & 26.6 (46.3) \dag& 9.90 (24.8)      & 6.30 (11.2)  & 0.90 (4.05) & \nodata (0.45) & \nodata (\nodata)& 17.1 (40.5)  \ddag \\
$\left[ SIII \right]$ 18.71$\mu$m    & 115. (199.) \dag& 36.4 (72.0)      & 18.4 (43.6)  & 7.65 (20.7) &  2.25   (7.20) & \nodata (0.45)   & 64.7 (144.)  \ddag \\
$\left[ OIV \right]$ 25.89$\mu$m     & 8.10 (15.3)     & 3.60 (9.90)     & 3.15 (11.2) & 0.90 (1.80) & \nodata (\nodata)& \nodata (\nodata)& 15.8 (38.2) \\
$\left[ SIII \right]$ 33.48$\mu$m    & 199. (326.)     & 180. (245.)      & 171. (270.)  & 71.1 (140.) &  31.5   (69.8)  & 5.85 (18.4)     & 658. (1069.) \\
$\left[ SiII \right]$ 34.81$\mu$m    & 256. (412.)     & 245. (329.)      & 270. (373.)  & 140. (240.) &  69.8   (136.)  & 18.4 (46.4)     & 999. (1536.) \\
$\left[ OIII \right]$  51.81$\mu$m   & 92.7 (219.)     & 180. (245.)      & 171. (321.)  & 48.1 (140.) &  4.05   (19.8)  & \nodata (0.45)  & 496. (945..) \\
$\left[ NIII \right]$ 57.32$\mu$m    & 153. (326.)     & 52.2 (122.)   & 18.4 (64.8)  & 0.45 (4.05) & \nodata (\nodata)& \nodata (\nodata) & 224. (517.) \\
$\left[ OI \right]$ 63.18$\mu$m      & 637. (939.)     & 329. (459.)   & 321. (422.)  & 140. (240.) &  31.5   (99.0)  & 5.85 (18.4) \dag   & 1458. (2159.) \ddag \\
$\left[ OIII \right]$  88.35$\mu$m   & 637. (939.)     & 212. (284.) & 171. (321.) & 71.1 (140.) \dag & 19.8 (48.1) \dag &1.35 (10.4) \dag & 1020. (1544.) \ddag \\
\enddata
\tablenotetext{*}{Notes: \#: total number of AGN inside detectable the SAFARI spectral range; 
\dag: outside the SAFARI spectral range;\ddag: excluding detections outside the SAFARI spectral range.}
\end{deluxetable}

\begin{deluxetable}{lccccccc}
\tabletypesize{\scriptsize}
\tablewidth{0pt}
\tablecaption{Total number of galaxies detectable in a SAFARI  survey of 0.5$^2$  
in IR lines as a function of redshift at 5$\sigma$ (3$\sigma$) in 1 hr.  integration per FoV, following \citet{val09}  
\label{tab:valiante} } 
\tablehead{ \colhead{line/redshift}& \colhead{0$<$z$<$0.75}  & \colhead{0.75$<$z$<$1.25}   & \colhead{1.25$<$z$<$1.75} & \colhead{1.75$<$z$<$2.25 }  & \colhead{2.25$<$z$<$2.75}  & \colhead{2.75$<$z$<$4}  &  \colhead{all redhifts \#} } 
\startdata
PAH(11.25$\mu$m)                   &3353 (5602) \dag & 2383. (4094.) \dag & 1074.  (2743.) \dag & 322. (676.) & 91.3 (308.)&  4.95 (19.8)& 418. (1004.)\ddag  \\
$\left[ NeII \right]$ 12.81$\mu$m  &1114 (1811) \dag &  251. (595.) \dag &  28.8  (136.) & 11.7 (49.1) & 1.8 (15.8)&\nodata (\nodata )  & 42.3 (201.)\ddag  \\
$\left[ NeIII \right]$ 15.55$\mu$m & 850 (1429) \dag &  94.0 (252.) &  51.8  (136.) & 26.1 (85.1) & 7.2 (29.2) &  \nodata  (4.95)	& 179. (507.)\ddag  \\
H$_2$(17.03$\mu$m)                 & 158 (331) \dag &  4.95 (17.1)& 1.35 (7.65) &\nodata (\nodata)&\nodata (\nodata) &\nodata (\nodata) & 6.3 (24.8)\ddag  \\
$\left[ SIII \right]$ 18.71$\mu$m  & 158 (331) \dag &  4.95 (17.1)& 1.35 (7.65) &\nodata (\nodata)&\nodata (1.80)&\nodata (\nodata)	& 6.3 (26.5)\ddag  \\
$\left[ OIV \right]$ 25.89$\mu$m   & 638 (1114) &  94.0 (252.) &  136.  (327.) & 49.1 (211.) & 15.8 (51.8)& \nodata (4.95)	        & 938. (1961.)\\
$\left[ SIII \right]$ 33.48$\mu$m  & 638 (1114) &  251. (879.) &  136.  (327.) & 49.1 (211.) & 29.3 (142.)& \nodata (4.95)	        & 1103. (2678.)\\
$\left[ SiII \right]$ 34.81$\mu$m  &1114 (1811) &  879. (1761.) &  505.  (1074.) & 322. (676.) & 207. (452.)&  9.9 (61.2)	        & 3037. (5835.)\\
$\left[ OIII \right]$ 51.81$\mu$m  &2258 (4010) &  879. (1761.) &  505.  (1512.) & 212. (676.) & 29.2 (142.)& \nodata (4.95)	        & 3883. (8106.) \\
$\left[ NIII \right]$ 57.32$\mu$m  & 850 (1811) &  94.0 (392.)&  28.8  (136.) &\nodata (11.7) &\nodata (\nodata) & \nodata (\nodata)	& 973. (2351.)\\
$\left[ OI \right]$ 63.18$\mu$m    &3353 (4762) & 3149. (5211.) & 1512.  (2743.) & 676. (1326.)& 207. (639.)& 19.8 (61.2) \dag        & 8897. (14681.) \ddag \\
$\left[ OIII \right]$ 88.35$\mu$m  &3353 (4762) & 1263. (2383.) &  505.  (1512.) & 322. (955.) \dag & 142. (307.)\dag &  4.9 (36.0)\dag & 5121. (8657) \ddag \\
\enddata
\tablenotetext{*}{Notes: \#: total number of AGN inside detectable the SAFARI spectral range; 
\dag: outside the SAFARI spectral range;\ddag: excluding detections outside the SAFARI spectral range.}
\end{deluxetable}

\begin{deluxetable}{lcccccc}
\tabletypesize{\scriptsize}
\tablewidth{0pt} 
\tablecaption{Number of galaxies detectable in a CCAT survey of 0.5$^2$  
in IR lines as a function of redshift at 5$\sigma$ (3$\sigma$) in 1 hr. integration per FoV, following \citet{fra10}  \label{tab:frances_ccat} } 
\tablehead{ \colhead{line/redshift}& \colhead{0$<$z$<$0.75}  & \colhead{0.75$<$z$<$1.25}   & \colhead{1.25$<$z$<$1.75} & \colhead{1.75$<$z$<$2.25 }  & \colhead{2.25$<$z$<$2.75}  & \colhead{2.75$<$z$<$4}  }
\startdata

$\left[ OI \right]$  63.18$\mu$m        & \nodata (\nodata)  & \nodata (\nodata)  & \nodata (\nodata) & 0.27 (3.60) $\diamond$ & \nodata (0.27) $\dag$ &  \nodata (0.22)$\ddag$\\
$\left[ OIII \right]$  88.35$\mu$m      & \nodata (\nodata)  & 22.2 (67.5) $\diamond$ & 1.80 (25.8) $\dag$ & 2.07 (12.1) $\ddag$ & 86.0 (219.) $\#$ & 21.1 (67.0) $\$$ \\
$\left[ NII \right]$  121.90$\mu$m      & 13.5 (39.0) $\dag$ & 0.22 (0.85) $\dag$  & 0.27 (1.80) $\ddag$ & 12.1 (59.0) $\#$& 3.15 (27.0) $\$$ &  0.22 (1.53) $\S$  \\
$\left[ OI \right]$  145.52$\mu$m      & 0.45 (1.35) $\dag$  & \nodata (0.09) $\ddag$  & 0.86 (3.02) $\#$ & \nodata (0.25) $\$$ & \nodata (\nodata) $\S$ & \nodata (0.72) $\P$  \\
$\left[ CII \right]$ 157.74$\mu$m      & 235. (643.) $\dag$  & 1786. (3951.) $\#$   & 1314. (1998.) $\$$     & 517. (1116.) $\$$ & 219. (603.) $\S$ & 350. (603.)   $\P$  \\
\enddata
\tablenotetext{*}{Notes: $\diamond$: at 200$\mu$m band;
$\dag$: at 230$\mu$m band; $\ddag$: at 291$\mu$m band; $\#$: at 350$\mu$m band; $\$$: at 450$\mu$m band; $\S$: at 620$\mu$m band; $\P$:at 740$\mu$m band}
\end{deluxetable}

\begin{deluxetable}{lcccccc}
\tabletypesize{\scriptsize}
\tablewidth{0pt} 
\tablecaption{Number of galaxies detectable in a CCAT survey of 0.5$^2$  
in IR lines as a function of redshift at 5$\sigma$ (3$\sigma$) in 1 hr. integration per FoV, following \citet{gru11}  \label{tab:gruppioni_ccat} } 
\tablehead{ \colhead{line/redshift}& \colhead{0$<$z$<$0.75}  & \colhead{0.75$<$z$<$1.25}   & \colhead{1.25$<$z$<$1.75} & \colhead{1.75$<$z$<$2.25 }  & \colhead{2.25$<$z$<$2.75}  & \colhead{2.75$<$z$<$4}  }
\startdata
$\left[ OI \right]$  63.18$\mu$m        & \nodata (\nodata)  & \nodata (\nodata)  & \nodata (\nodata) & 2.02 (13.7) $\diamond$ & 0.36 (1.53) $\dag$ &  0.22 (1.71)  $\ddag$ \\
$\left[ OIII \right]$  88.35$\mu$m      & \nodata (\nodata)  & 22.1 (53.5) $\diamond$ & 5.49 (28.6) $\dag$ & 7.47 (25.3) $\ddag$ & 108. (240.) $\#$ & 39.7 (118.) $\$$ \\
$\left[ NII \right]$  121.90$\mu$m      &12.7 (32.5) $\dag$ & 0.81 (3.01) $\dag$  & 1.53 (5.49 $\ddag$ & 25.3 (71.1) $\#$& 11.0 (37.9) $\S$ &  1.71 (6.12) $\S$  \\
$\left[ OI \right]$  145.52$\mu$m      & 0.81 (2.02) $\dag$  & \nodata (0.45) $\ddag$  & 2.65 (9.85) $\#$ & 1.00 (3.60) $\$$ & 0.13 (0.72) $\S$ & 0.49 (3.01) $\P$  \\
$\left[ CII \right]$ 157.74$\mu$m      & 172. (432.) $\dag$  & 1728. (3262.) $\#$   & 1498. (2628.) $\$$     & 473. (1089.) $\$$ & 240. (534.) $\S$ & 572.  (882.)   $\P$  \\
\enddata
\tablenotetext{*}{Notes: $\diamond$: at 200$\mu$m band;
$\dag$: at 230$\mu$m band; $\ddag$: at 291$\mu$m band; $\#$: at 350$\mu$m band; $\$$: at 450$\mu$m band; $\S$: at 620$\mu$m band; $\P$:at 740$\mu$m band}
\end{deluxetable}

\begin{deluxetable}{lcccccc}
\tabletypesize{\scriptsize}
\tablewidth{0pt} 
\tablecaption{Number of galaxies detectable in a CCAT survey of 0.5$^2$  
in IR lines as a function of redshift at 5$\sigma$ (3$\sigma$) in 1 hr. integration per FoV, following \citet{val09}  \label{tab:valiante_ccat} } 
\tablehead{ \colhead{line/redshift}& \colhead{0$<$z$<$0.75}  & \colhead{0.75$<$z$<$1.25}   & \colhead{1.25$<$z$<$1.75} & \colhead{1.75$<$z$<$2.25 }  & \colhead{2.25$<$z$<$2.75}  & \colhead{2.75$<$z$<$4}  }
\startdata
$\left[ OI \right]$  63.18$\mu$m        & \nodata (\nodata)  & \nodata (\nodata)  & \nodata (\nodata) & \nodata (26.4) $\diamond$ & \nodata (1.94) $\dag$ &  \nodata (0.09)  $\ddag$ \\
$\left[ OIII \right]$  88.35$\mu$m      & \nodata (\nodata)  & 17.5 (54.9) $\diamond$ & 3.69 (28.7) $\dag$ & 12.0 (49.5) $\ddag$ & 208. (450.) $\#$ & 36.1 (97.6) $\$$ \\
$\left[ NII \right]$  121.90$\mu$m      &13.1 (40.5) $\dag$ & 0.13 (1.00 $\dag$  & 0.13 (3.69) $\ddag$ & 49.5 (136.) $\#$ & 29.2 (91.3) $\$$ &  0.09 (4.81) $\S$  \\
$\left[ OI \right]$  145.52$\mu$m      & 0.94 (1.44) $\dag$  & \nodata (0.09) $\ddag$  & 1.35 (7.65) $\#$ & \nodata (3.73) $\$$ & \nodata (\nodata) $\S$ & \nodata (1.66) $\P$  \\
$\left[ CII \right]$ 157.74$\mu$m      & 231. (639.) $\dag$  & 2385. (5211.) $\#$   & 2061. (3586.) $\$$     & 954. (2371.) $\$$ & 450. (1215.) $\S$ & 531. (1075.)   $\P$  \\
\enddata
\tablenotetext{*}{Notes: $\diamond$: at 200$\mu$m band;
$\dag$: at 230$\mu$m band; $\ddag$: at 291$\mu$m band; $\#$: at 350$\mu$m band; $\$$: at 450$\mu$m band; $\S$: at 620$\mu$m band; $\P$:at 740$\mu$m band}
\end{deluxetable}

\end{document}